\def\preprint{1}		
\def\comment#1{}

\if\preprint1
	\documentclass[usenatbib]{mn2e}
        \usepackage{natbib}
	\usepackage{times}
	\usepackage{graphicx}
	\usepackage{calc}
         \usepackage{subfig}
         \usepackage{color}
\else
	\documentstyle[astrop-bib,referee,times]{mn}
	\newcommand{\includegraphics}[1]{}
\fi

\def\oversim#1#2{\lower0.5pt\vbox{\baselineskip0pt \lineskip-0.5pt
     \ialign{$\mathsurround0pt #1\hfil##\hfil$\crcr#2\crcr\sim\crcr}}}

\title[The nuclear PAH emission of merger system NGC 1614]
{The nuclear PAH emission of merger system NGC 1614: rings within rings }
\author[P. V\"ais\"anen et al.]
       {Petri V\"ais\"anen$^{1,2}$ \thanks{E-mail: \tt petri@saao.ac.za}, Vinesh Rajpaul$^3$, Albert~A.~Zijlstra$^4$,   Juha Reunanen$^5$,  Jari Kotilainen$^6$
        \\
       $^1$South African Astronomical Observatory, P.O. Box 9, Observatory 7935, Cape Town, South Africa,\\	
       $^2$Southern African Large Telescope, P.O. Box 9, Observatory 7935, Cape Town, South Africa,\\	
       $^3$Department of Astronomy, University of Cape Town, Private Bag X3, Rondebosch 7701, South Africa\\
       $^4$Jodrell Bank Centre for Astrophysics, School of Physics and Astronomy, University of Manchester, Oxford Road, Manchester M13 9PL, UK\\
      $^5$Tuorla Observatory, University of Turku, FI-21500 Piikki\"o, Finland\\
      $^6$Finnish Centre for Astronomy with ESO (FINCA), University of Turku, V\"ais\"al\"antie 20, FI-21500 Piikki\"o, Finland
}
\pubyear{2011}

\begin{document}

\maketitle

\begin{abstract}

It is important to understand the interplay between nuclear star-formation and nuclear activity when
studying the evolution of gas rich galaxy mergers.  We present here new spatially resolved $L$-band 
integral field unit observations of the inner kpc of the luminous IR galaxy NGC~1614. 
A broad ring of 3.3 $\mu$m PAH emission is found at a distance of approximately 200 pc from the core. 
This ring overlaps with a previously established star-forming ring detected with Pa$\alpha$ and radio
continuum observations, but peaks outside it, especially if determined using the PAH equivalent width.
Using the characteristics of the PAH emission and the ionised gas emission we argue that NGC~1614
features an outward propagating ring of star formation, 
where the equivalent width of the PAH emission localises
the regions where the current star formation is just expanding into the molecular gas outward of the nucleus. 
The core itself shows a highly luminous, slightly resolved (at $\sim80$ pc) $L$-band continuum source.
We find no evidence of AGN activity and rule out the presence of an obscured AGN using $L$-band
diagnostics.  Furthermore, we detect the likely companion galaxy from archival {\em HST}/ACS 
imaging. The star formation and dynamical characteristics of the system are consistent with a relatively 
major merger just after its second passage.  An outstanding question is how a gas-rich 
advanced merger such as this one, with strong LIRG level nuclear starburst and major merger-like 
tidal features, has not yet developed an active nucleus.

\end{abstract}

\begin{keywords}
galaxies: nuclei -- galaxies: starburst -- galaxies: active  -- galaxies: NGC 1614
\end{keywords}

\section{Introduction}

In the local Universe, luminous and ultraluminous infrared galaxies, 
(U)LIRGs\footnote{ULIRGs are defined as 
$\log L/L_{\odot}>12.00$, and LIRGs as $\log L/L_{\odot}=11.00-11.99$}, 
are strongly related to  merging activity.  In a traditional view 
\citep[see e.g.][] {Sanders1988,Barnes1992,Hopkins2006,Lonsdale2006}, 
a merger between two spiral 
galaxies initiates with a  close approach, with a second close 
approach on the order of $10^8$ yr later, 
lasting on the order of $10^7$ yr, followed by rapid coalescence, where the 
time scales are parameter-dependent.  A starburst first occurs during the
early stages. At later stages gas and dust are efficiently channeled into the
nuclear regions, leading to both starburst and strong AGN activity. Feedback
from the massive energy production in the heavily obscured AGN will
subsequently quench star formation \citep[e.g.][]{Springel2005,Johansson2009b} 
and drive out the remaining interstellar matter, revealing an optical QSO and eventually  
an elliptical galaxy with a  'dead'  supermassive black-hole. 
The starburst activity, and thus the (U)LIRG
phenomenon, is likely to occur episodically during both early and later phases
of a merger \citep{Murphy2001,Hopkins2006}.
(U)LIRG-level star formation appears to begin while the progenitors are still separately
distinguishable \citep{Murphy1996} and a large fraction of them shows evidence 
for ongoing merging \citep[][]{Farrah2001, Surace2000}, often with compact emission 
knots suggesting the formation of super star clusters. An important question is what triggers the star
formation and how the (circum)nuclear star formation interacts with the
nuclear activity \citep[e.g.][]{Laine2006,Watabe2008}.

Whatever the details, to understand and establish the transformations 
of spirals into QSOs via (U)LIRGs, it is crucial to investigate 'composite' 
stages where both starbursts and AGN potentially co-exist 
\citep[e.g.][]{Yuan2010}.   It is important to {\em localise} the 
starburst activity as well as to locate the early, buried AGN activity, 
in order to understand the triggering processes and any causal relations. 
During the coalescence, star formation (SF) is expected in the
nuclear region where the gas density is highest, though more spread out
SF is possible as well \citep{Alonso2006}. 
The required high-resolution observations are
complicated for two reasons. Firstly, optical extinction is always high and
major parts of the system are detectable only in the infrared. Even HST images
can miss complete galaxies and major nuclei \citep{Vaisanen2008,Haan2011}. 
Secondly, the important
mid-infrared Spitzer images showing the location of heated dust have poor
angular resolution.  Clarifying the environment and separating the various
AGN and starburst components requires high-resolution infrared observations.

The little explored 3--4 $\mu$m $L$-band in the mid-infrared is promising for
separation of AGN and starbursts. The $L$-band continuum would be dominated
by an AGN even if its bolometric luminosity would be minor; a starburst would
yield strong 3.3 $\mu$m PAH emission, and cool dense gas around an AGN can show a 
3.4 $\mu$m absorption feature \citep[see e.g.][]{Risaliti2006,Sani2008}.
The pioneering $L$-band spectroscopic surveys of starburst galaxies reported
in \citet{Imanishi2000,Imanishi2006, Risaliti2006} made use of long-slit spectra.
Studying the spatial distribution of PAH within the nuclear regions of starbursts 
was not possible in those observations and has remained 
challenging due to the required angular resolution.
Among the few studies to date, \citet{Tacconi2005} found from high resolution imaging 
that the 3.3 $\mu$m PAH emission globally peaks on central starburst regions, 
but on smaller scales the correlation is not clear. Other ground-based studies 
have used PAH features at $\sim8.8$ $\mu$m or $\sim11$  $\mu$m to localise starbursts 
and nuclear activity \citep{Diazsantos2008,Siebenmorgen2008}. 
Resolving and localising the 3.3 $\mu$m PAH emission from 
other SF indicators, such as radio and Pa$\alpha$, and peaks of mid-infrared (MIR) 
continuum, is important to be able to minimise uncertainties regarding whether PAHs 
are destroyed only by AGN or also by young SF regions 
\cite[see e.g.][]{Mason2007}. The detailed relationship between circumnuclear star
formation and PAH emission is still unclear. 

Integral field spectroscopy provides significant advantages especially for
disturbed, compact systems, providing spatially
resolved spectral coverage over a limited field of view. Long-slit
observations, in contrast, will miss any emission outside of the narrow
slit, while the use of narrow-band imaging is limited by the fact that there are often 
no suitable filters in existence for redshifted PAH features.  
The power of IR integral field unit (IFU) observations for ULIRGs has been
demonstrated by e.g.\ \citet{Reunanen2007} and \citet{Arribas2008}, 
but so far only at $H$- and $K$-bands.  
In this paper, we present new $L$-band IFU spectroscopy of the 
nucleus of the LIRG NGC~1614.  NGC~1614 is considered a `laboratory' 
\citep{Alonso2001} for studying nuclear starbursts with its extremely strong nuclear 
IR emission and signs of recent interactions and advanced merging activity,
as well as possible nuclear activity.  It is an ideal target to study the physical details
of the gas-rich galaxies transforming into QSOs and ellipticals via (U)LIRGs and 
obscured AGN.  Our data allow us to study the spatial distribution of the PAH emission,
to investigate its relation to the SF regions, and to separate out potential contributions from 
an AGN nucleus at high spatial resolution.

\section{NGC 1614}

\begin{figure*}
\includegraphics[width=\textwidth,clip=true]{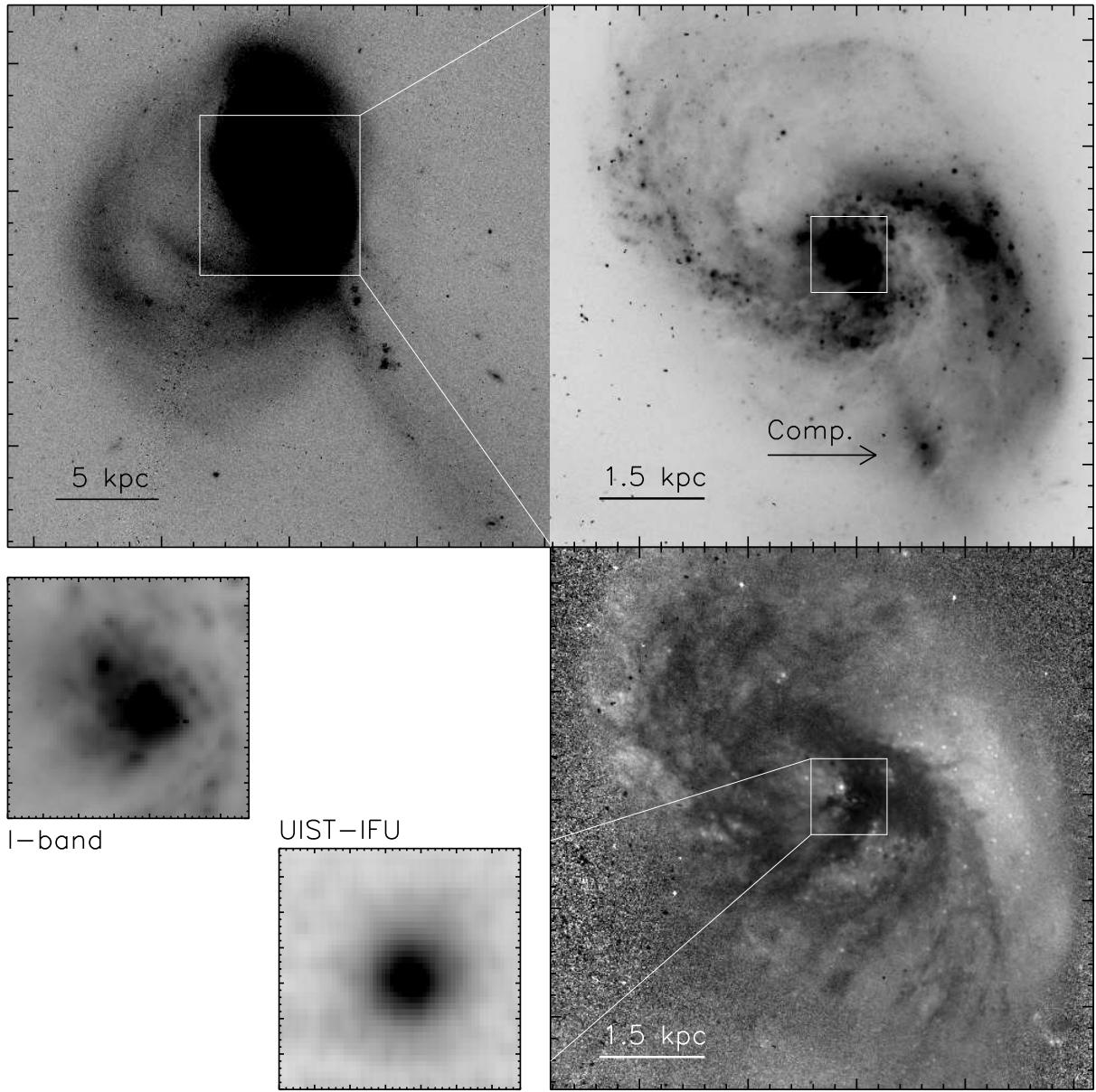}
\caption{\label{hstpics}  Archival ACS/HST images of NGC~1614. North is up and 
East left in each image, the tick-marks are 5\arcsec\ in the top-left image
and 1\arcsec\ in the right side panels.  {\em Top-left}:  an $I$-band (FW814) image 
showing the outer structures of NGC~1614 at 20 kpc scales. The white rectangle 
shows the area of the same image zoomed and re-scaled at {\em top-right}, 
showing the more regular spiral pattern. The higher surface density structure
9\arcsec\ SW of the nucleus is pointed out, which is the likely remnant of a companion
galaxy (Section~\ref{companion}).  The {\em bottom-right} panel shows 
the ACS/HST $B-I$ colour map of the same region as the panel above it, where 
darker areas mean redder colours, i.e.\ higher extinction.  The main nucleus is 
seen to lie behind a dust feature, while there is a very blue point 
source 1\arcsec\ NE of the nucleus (see Section~\ref{secondnucleus}). 
The innermost 1 kpc area ($3.4\arcsec \times 3.4\arcsec$) is marked with the white
rectangles in both right-hand-side panels and is shown zoomed-in at  {\em bottom-left}.
The left one is the $I$-band image, and at right we show our UIST
$L$-band continuum image of the nuclear region using the same spatial scale.}
\end{figure*}

NGC~1614 is a luminous infrared galaxy with an infrared luminosity is $\log L/L_{\odot} = 11.6$ 
at a distance of 64 Mpc (redshift $z=0.016$).   Optically it is classified 
as a border-line case
between a LINER and HII-dominated galaxy \citep{Bushouse1986,Veilleux1995}, and \citet{Yuan2010} classify it as a composite starburst + AGN.   
Morphologically the galaxy is barred and interacting 
\citep[e.g.][]{deVauc1976,Neff1990}; 
at large scales it shows strong perturbations indicative of an
earlier major interaction, though the appearance has been attributed to a 
lower-mass galaxy too \citep{Neff1990}.  
While it is optically quite chaotic with long curved
asymmetric tail(s) to the East and a 22~kpc linear structure towards South-West, 
it has two clear inner spiral arms (Fig.~\ref{hstpics}).  

The central region of NGC~1614 
was studied by \citet{Heisler1999} and \citet{Kotilainen2001} 
using Br$\gamma$ and H$_2$ observations at AAT and UKIRT, respectively,
and by \citet{Alonso2001} using Pa$\alpha$ emission with the HST/NICMOS.
All found a star-forming ring of $\sim600$~pc diameter around the nucleus.
Recently \citet{Olsson2010} studied the starburst with multi-wavelength sub-mm
and radio data.  None of these studies found clear evidence of AGN activity in
NGC~1614 and the star formation rate (SFR) was estimated to be $\sim 50$ M$_{\odot}$ yr$^{-1}$.

$H$ and $K$-band observations have been presented by \citet{Puxley1999} and Spitzer
spectroscopy is presented by \citet{Bernard2009}. An AKARI 2.5--5 $\mu$m
spectrum was recently published by \citet{Imanishi2010}.  In addition, a  low-resolution 
$L$-band spectrum of NGC1614 was published by \citet{Mizutani1994}.  
Figure~\ref{space} shows the part of the
Spitzer IRS spectrum covering the PAH bands: these data were retrieved from
the Spitzer archive and pipeline reduced.
The insert shows the AKARI
spectrum of \citet{Imanishi2010}.

\section{Observations}

We used UIST \citep{Ramsay2004}, a 1--5 $\mu$m imager-spectrometer
with a $1024 \times 1024$ InSb array, at UKIRT.  Integral field unit (IFU) spectroscopy
in the $L$-band was an observing mode unique to UIST; it is not available anywhere
else currently.  IFU observations with UIST
use an image slicing design. The image slicing mirror comprises 18 segments
or  'slices' that are each $0.24\arcsec \times 6.00\arcsec$ in size; 14 adjacent slices yield a
spatial coverage on the sky of $3.3\arcsec \times 6.0\arcsec$, with a scale of 
$0.12\arcsec \times 0.24\arcsec$ per pixel. Images ('channels') can be extracted for
every resolution element of the grism.

The observations were carried out over two nights, 2009 January 6 and 26.  
We used the short-$L$ grism, which provided a resolution of 700 and a wavelength coverage of
2.905--3.638 $\mu$m.  Total observing time for the IFU observations was 160
min, half of which was spent on-source and half on-sky; the observations
alternated between object and sky observation every 60 sec. The second night was
affected by some cloud, and an additional sky subtraction using the outer
region of the IFU field was required.  After this additional step, the two
observations agreed very well.  

Data reduction was done using ORAC-DR \citep{Cavanagh2008}, with a
pipeline reduction developed at the Joint Astronomy Centre, resulting in 
wavelength and flux calibrated datacubes.  Wavelength calibration was done 
using an arc as there are no suitable telluric emission lines in this wavelength
range.  A standard star, BS1543, was observed and used for relative flux calibration.
For an absolute flux calibration the resulting datacubes were scaled to broadly agree with 
the AKARI NGC~1614 spectrum. This constant scaling (reflected in the numerical scales of
several Figures in this paper) does not in any way affect our results 
which are all based on relative fluxes and their spatial distribution and equivalent 
widths of spectral features.
In addition, an automated  script was implemented to perform robust identification and 
removal of some nonphysical `artefact' pixels that were present in the datacubes;
the problematic fluxes were replaced with values interpolated from 
contiguous pixels. The separate
datacubes from the two different observations were then aligned spatially, 
at sub-pixel resolution, and co-added to yield a single datacube
on which all subsequent analyses were carried out.

We also obtained imaging in two filters, nbL (3.379--3.451 $\mu$m, overlapping
with the grism coverage) and $L^\prime$ (3.428--4.108 $\mu$m, longward of the
spectral coverage). Total exposure times were a few minutes per filter,
yielding shallow images only. At the $0.12\arcsec$ plate scale, total field of
view is about 1 arcmin. We do not use these data further, except to note that the
nucleus is detected in both images, and clearly is the only significant 
source of $L$-band emission in NGC~1614. This is confirmed from archival Spitzer imaging
as well, where we measure nearly half of the total NGC~1614 broad band 3.6 $\mu$m flux 
as originating from inside a 2\arcsec\ aperture radius.

\section{Results}

\subsection{The nuclear source and continuum flux}

\begin{figure}
 \includegraphics[width=9cm,clip=true]{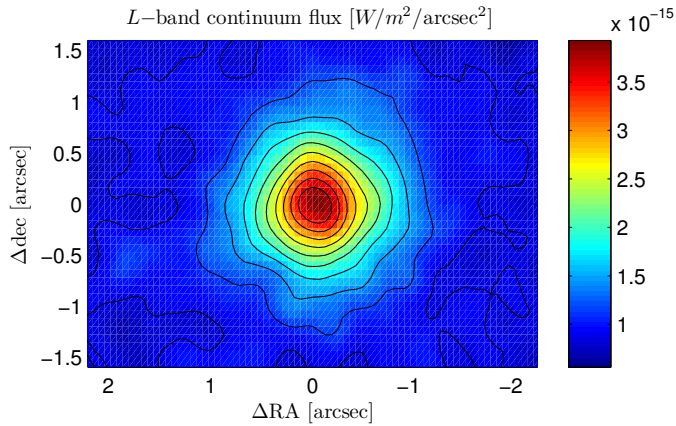}
 \caption{\label{contmap}  The $L$-band continuum map of NGC~1614: the integrated
 UIST image obtained by summing over all channels excluding those around
 the PAH emission features. 
}
 \end{figure}

High-resolution images of the central area were obtained from the IFU data, by
summing over selected channels from the spectroscopic datacube.
The continuum image in Fig.~\ref{contmap} was created by summing all 1024 channels, 
except for channels 600-700 and 750-850 corresponding to the redshifted 3.3 and 
3.4 $\mu$m features: it confirms the dominance of a
central source.  The bright source has a FWHM of 0.42 arcsec.  The seeing was determined
from the IFU data of the standard star: a
FWHM of 0.33 arcsec was obtained. This indicates the central source is
marginally resolved, with a deconvolved FWHM of 0.26 arcsec, or 80 pc.
The HST NICMOS $J$- and $H$-band images \citep{Alonso2001} show a more compact
central source, slightly resolved at 0.10\arcsec\ deconvolved.

Our IFU data $L$-band continuum source is defined as the nucleus of the galaxy here, 
and all maps and radial plots henceforth are with respect to this point. Also, when 
comparing to other data, as for example the NICMOS data above, maps are aligned using the
position of the continuum source.  

\subsection{The spectrum and maps of the central region}
\label{ifuspectra}

\begin{figure}
  \begin{minipage}[t]{150pt}
    \includegraphics[width=8cm,clip=true]{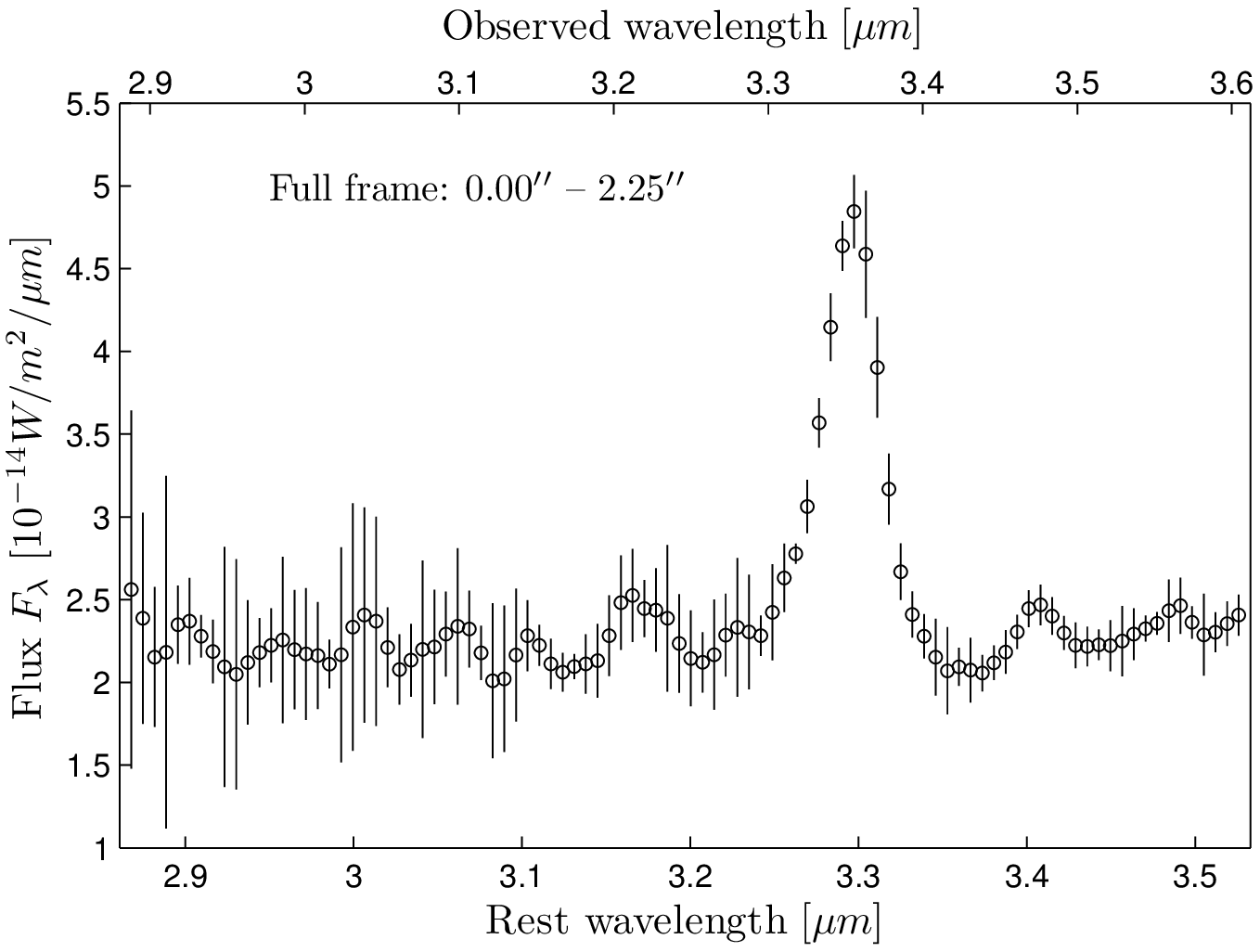}
  \end{minipage}
  \hfill
  \begin{minipage}[t]{150pt}
    \includegraphics[width=8cm,clip=true]{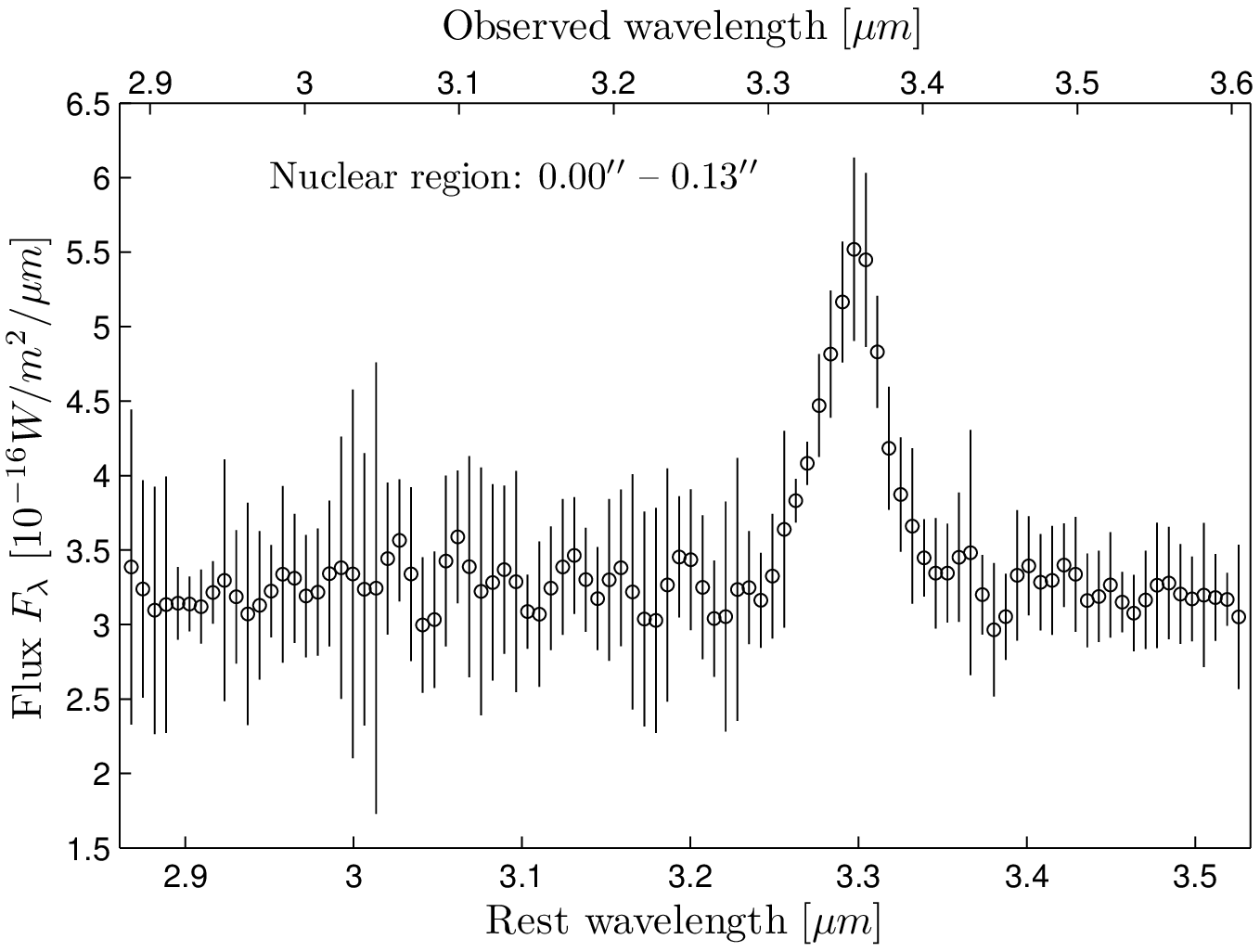}
  \end{minipage}
  \hfill
   \begin{minipage}[t]{150pt}
    \includegraphics[width=8cm,clip=true]{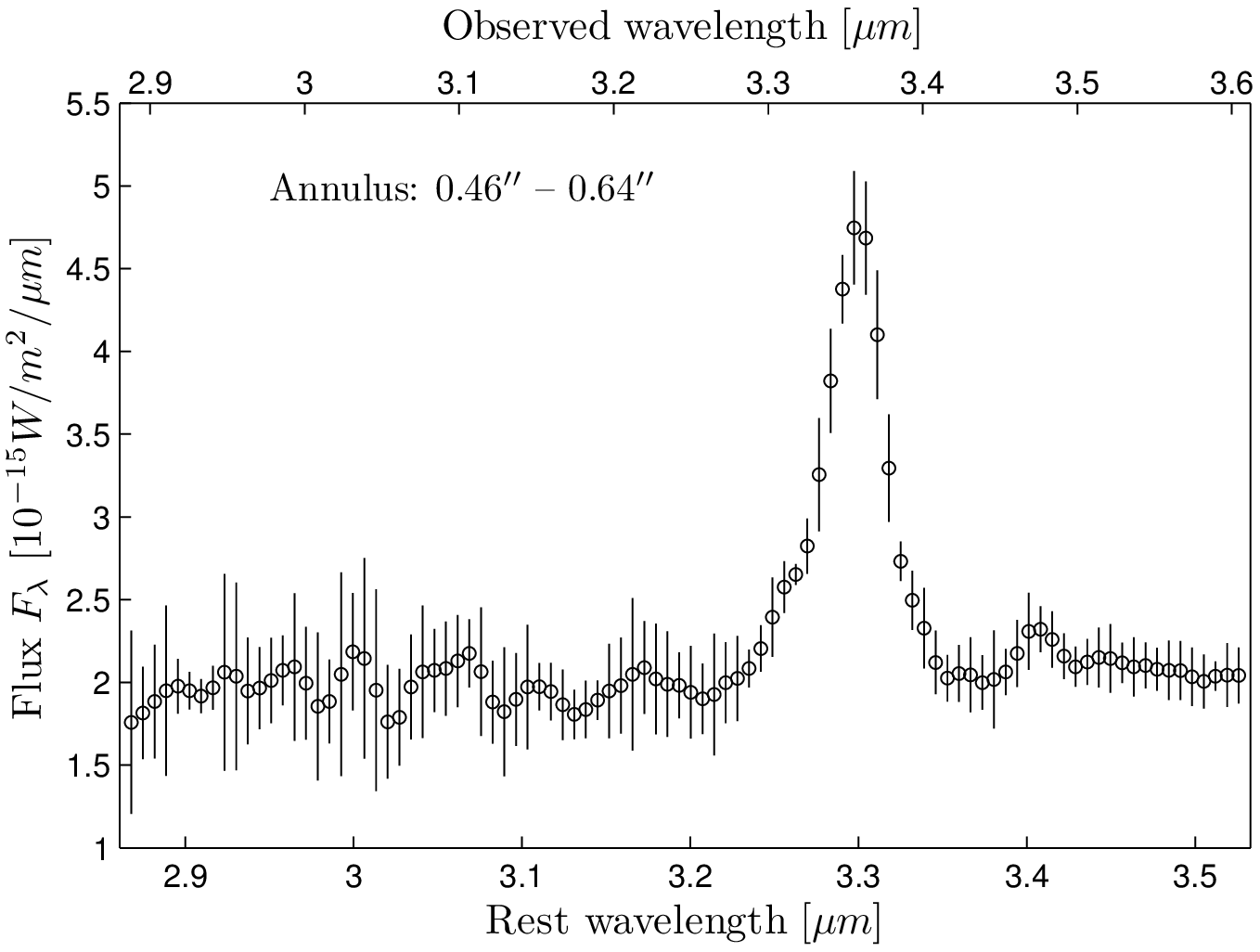}
  \end{minipage}
  \caption{\label{spec}  Top: The integrated spectrum over  
a circular region with radius 2.25 arcsec centred on the continuum 
source.   Middle: The same, but corresponding only to the inner $r=0.13\arcsec$ region.  
Bottom: The same, but integrated over an annulus corresponding to the strongest PAH emission.}
\end{figure}

Spectra of different observed regions of NGC~1614 were obtained from the IFU data by integrating
each channel within specified apertures or annuli.
The spectrum of the central region a circular aperture of 2.25\arcsec\ radius
is shown in the top panel of Fig.~\ref{spec}.  A prominent 3.3 $\mu$m PAH feature
(C-H stretching vibration), a weaker 3.4  $\mu$m PAH band, and an otherwise flat continuum,
are evident.  The middle panel of  Fig.~\ref{spec} shows the IFU spectrum summed up in 
in the innermost nuclear region only, and the bottom panel in an annulus around the nucleus.
The notable difference is in the relative strength of the PAH features, though it is significant that there
still nevertheless is unmistakable PAH emission right in the core of NGC~1614, within the central 
80 pc (middle panel), in the smallest independent spatial resolution element of our data.

The Spitzer and AKARI  spectra are shown for comparison in Fig.~\ref{space}.  
They show strong PAH bands, superposed on a dust continuum longward of 10 $\mu$m, 
but a flat continuum shortward of 6 $\mu$m.  Hydrogen emission lines, especially Br$\gamma$ 
at 4.08 $\mu$m, and [Ar{\sc ii}] 6.99 $\mu$m are also visible.
The space-based spectra were obtained in
larger apertures than our IFU area;  the slit width is 3.7\arcsec\ for the shortest wavelengths 
in the Spitzer data while the AKARI spectrum is taken with an unspecified aperture from
{\em slit-less} spectroscopy in a $1\arcmin \times 1\arcmin$ field of view \citep[see][]{Imanishi2010}. 

The IFU spectrum has considerably higher spectral resolution than the AKARI data.
Comparing the two spectra shows general agreement in shape, 
but for one notable difference.
In the AKARI spectrum (Fig.~\ref{space} inset), there is a broad plateau redward of the PAH feature,
stretching from 3.35 to 3.45 $\mu$m. In the IFU spectrum (Fig.~\ref{spec}, top panel), 
the plateau is absent and instead there is a narrower band between 3.37 and 3.42 $\mu$m, well
separated from the PAH feature. The difference is most likely due to the
difference in spectral resolution, the gap between the bands being lost at the
resolution of AKARI. 
The line-to-continuum ratio at 3.3$\mu$m is similar, $\sim 2.2$, 
in both spectra in spite of the lower resolution of the AKARI spectrum. 
This might indicate that the latter may include some additional PAH emission within its 
much larger aperture. 

The IFU data were used to create separate image maps for the 3.3 and 3.4 $\mu$m 
PAH bands. The continuum was subtracted from these 
to produce band-only images, and finally the band-only images were divided by the 
continuum in order to also obtain equivalent-width (EW) maps.  
The 3.3 $\mu$m PAH flux map in Fig.\ref{PAH12}, top panel,
shows the uneven spatial distribution of the PAH emission.
The lower panels show the resulting EW maps for both bands;
in the case of the 3.4 $\mu$m feature, the PAH emission is adequately 
detectable only in EW.
These maps clearly trace the strongest PAH emission in a ring-like structure 
around the nucleus, which itself is strongly continuum dominated (Fig. \ref{contmap}).

Defining, arbitrarily, the PAH ring by the points where the 3.3~$\mu$m 
EW is within 20 per cent the maximum, the inner radius is at about 0.6\arcsec\ radial
distance from the nucleus, and has an angular thickness of 1.1\arcsec. These
translate to inner and outer radii of 200 and 550 pc, with the peak of the 
EW at approximately 250 pc radius.

\subsection{Radial dependencies and comparison with a star-formation ring}
\label{radialdeps}

\begin{figure*}

\includegraphics[width=8cm,clip=true]{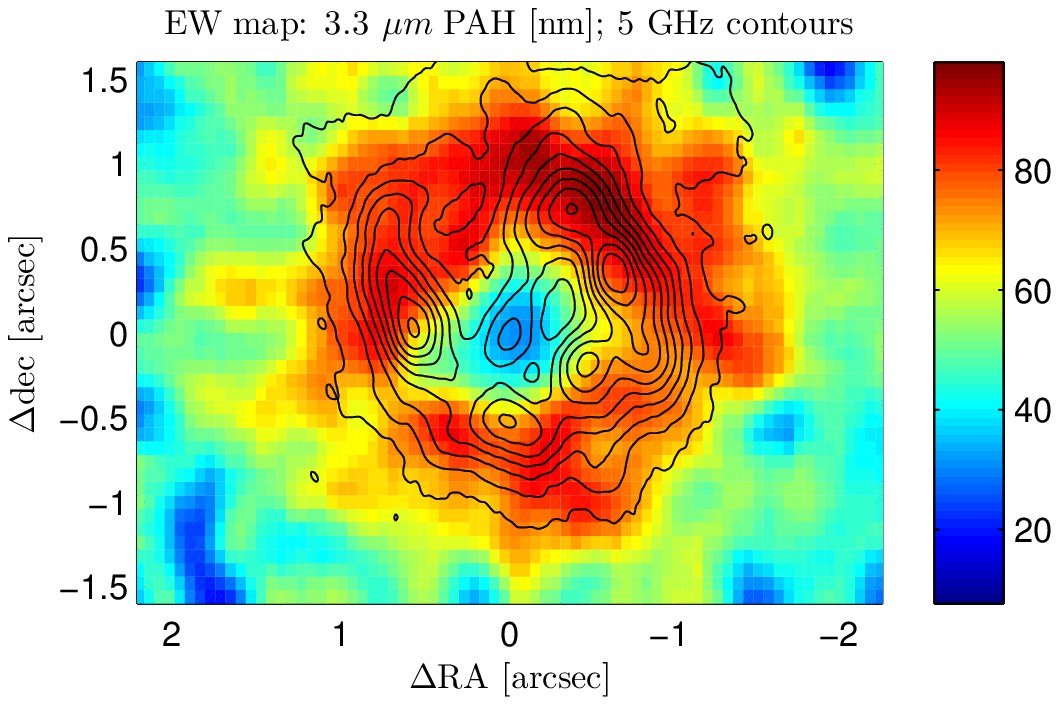}
\hspace{5mm}
\includegraphics[width=8cm,clip=true]{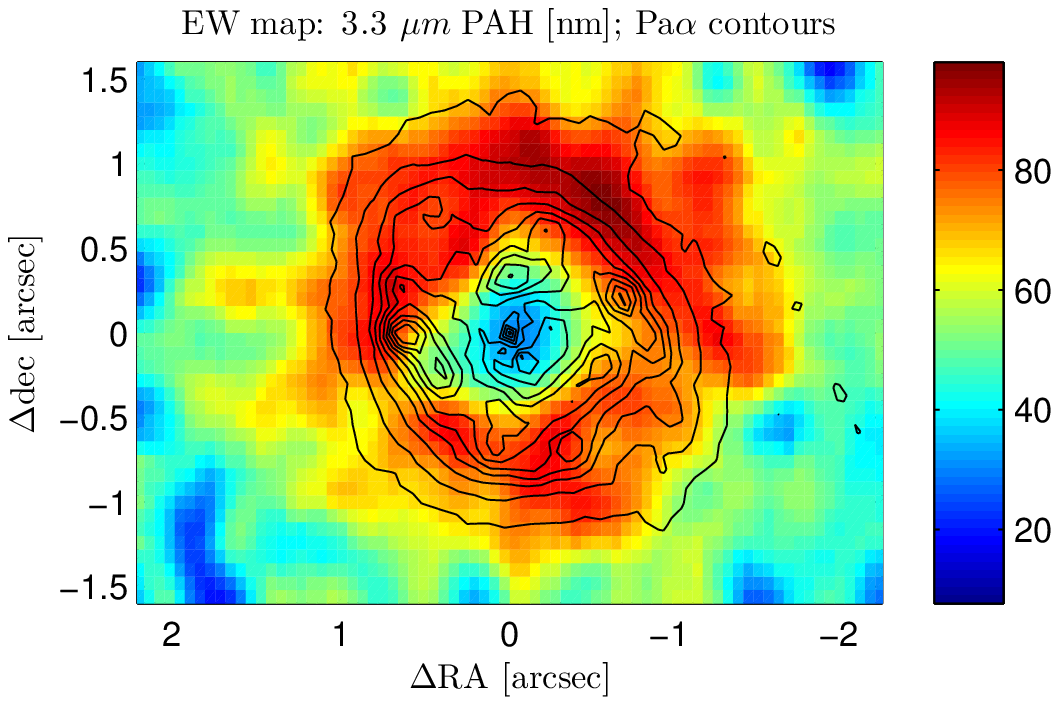}
\caption{\label{sfring1}
The  radio and Pa$\alpha$ emission is shown with linearly spaced contours overlaid on the 3.3 $\mu$m PAH EW map.
}
\end{figure*}

Figure~\ref{sfring1} shows the 3.3 $\mu$m PAH EW over-plotted with archival 
NICMOS Pa$\alpha$ emission \citep{Alonso2001} and with radio continuum of 
the NGC~1614 nuclear regions
from \citet{Olsson2010}.  The radio continuum and the Pa$\alpha$ trace each
other remarkably well, while the PAH EW appears to peak slightly further out. 
If the contours are over-plotted on the PAH {\em flux}, the peaks of the 
distributions are closer (Fig.~\ref{sfring2}).  The maps also suggest  the weakest PAH flux at the 
opening of the horse-shoe shape South-East of the centre might correlate with the
strongest Pa$\alpha$ and to a lesser extent with the radio emission.

To quantify the appearance of the different radial distributions of the 'rings',
Fig.~\ref{radplot} shows the azimuthally averaged 3.3$\mu$m PAH EW as well as the 
3.3 PAH flux as a function of radial distance from the core. We also plot the radial dependence of the 
Pa$\alpha$ flux measured from the archival NICMOS image showing an identical distribution
that presented in \citet[][]{Alonso2001}. The PAH flux and the Pa$\alpha$ peak at 0.5\arcsec\ (150 pc) 
while the PAH EW peaks at $\sim$0.8\arcsec\ (250 pc) from the core.   Note also that there 
is significant PAH flux in the core regions, while it weakens there in terms of 
equivalent width, and the Pa$\alpha$ shows a definite hole in the middle. The spatial resolution of the data are slightly different, but even convolving the Pa$\alpha$ distribution with
the UIST 0.3\arcsec\ PSF does not change the result.

\subsection{The PAH features}

\begin{figure}
\includegraphics[width=8.6cm,clip=true]{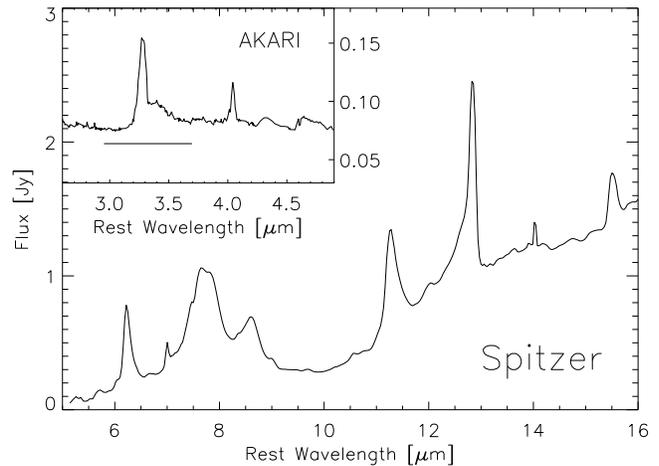}
\caption{\label{space}  Infrared spectra of
NGC~1614.  The large panel shows the Spitzer IRS spectrum, dominated by PAH
bands.  The inset shows the AKARI spectrum \citep[from][]{Imanishi2010}, with
a strong 3.3$\mu$m PAH band, and a 3.4$\mu$m wing. The wavelengths are in the rest
frame (redshift $z=0.016$). 
The horizontal bar below the AKARI spectrum shows the spectral range of the IFU observations 
in this paper (corrected for redshift). 
}
\end{figure}

PAH molecules show features over a range of wavelengths from 3.3 to 19
$\mu$m.  The longer-wavelength bands are covered by Spitzer (Fig.~\ref{space}) 
extracted using a larger aperture.  The UKIRT
observations indicate that only the core region contributed significantly and we 
will therefore assume that the UKIRT and Spitzer PAH bands  arise in the same location.

A detailed discussion of PAH emission can be found in \citet{Tielens2008}. PAH
emission sources are classified into three classes, based on wavelengths
especially of the C-C bands. 
Class A corresponds to PAH bands seen in the ISM, in gas illuminated by nearby
stars or ionised regions, while class B is seen in circumstellar
material, with class A possibly occuring at hotter temperatures 
\citep{Keller2008}. 
The short wavelength peak at 6.2 $\mu$m in NGC~1614 indicates class A PAH.  
However, the 8 $\mu$m band in NGC~1614 shows two equal peaks
at the class A and B wavelengths.
The PAH emission in NGC~1614 thus traces ISM gas in a relatively hot radiation field but there may
also be a component in a more subdued radiation field.
Notable is the strong 12.7 $\mu$m band which exceeds the 11.2 $\mu$m band strength.
This is a characteristic of ISM PAHs which are more irregular/ragged,
and indicative of regions of massive star formation.
The 6.2/11.2 band strength ratio indicates that the PAHs are mostly neutral
\citep{Galliano2008}, 
and this is also indicated by the strength of the
12.7 $\mu$m band.   In fact, neutral PAHs are much stronger at 3.3 $\mu$m
than are ionised PAH, and the UKIRT observations would be insensitive to
ionised PAHs.  The 3.4 $\mu$m band is attributed to an aliphatic C-H stretch, which may be
related to partial PAH destruction.

\begin{figure}
  \begin{minipage}[t]{150pt}
    \includegraphics[width=8.45cm,clip=true]{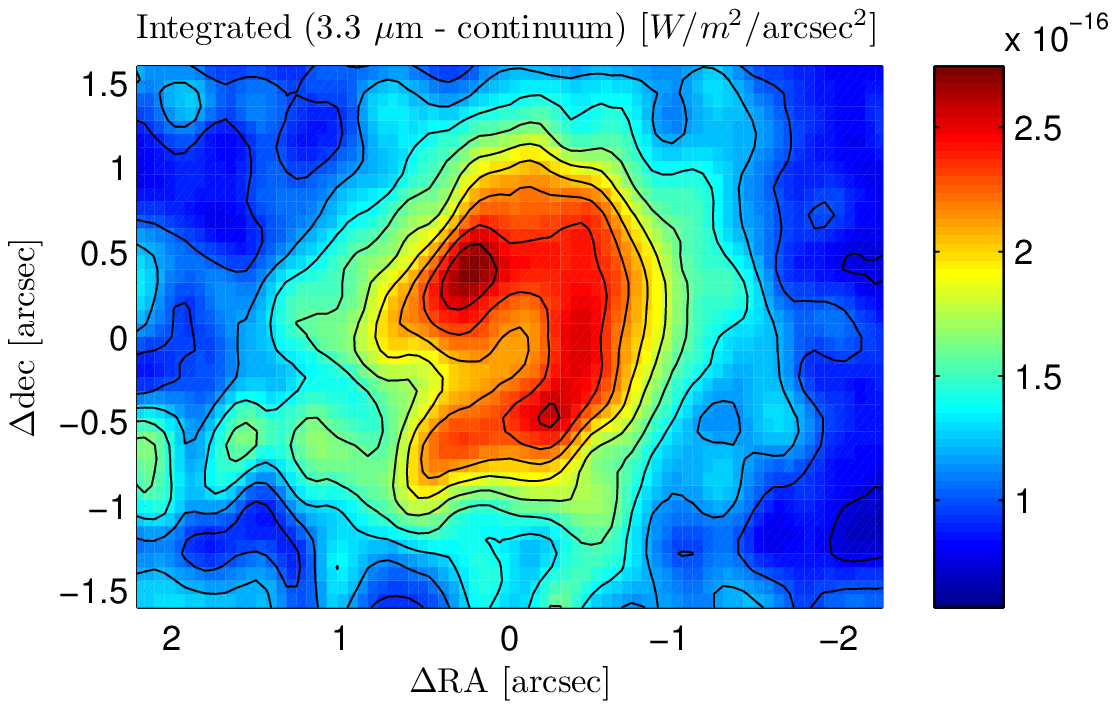}
  \end{minipage}
  \hfill
  \begin{minipage}[t]{150pt}
    \includegraphics[width=8cm,clip=true]{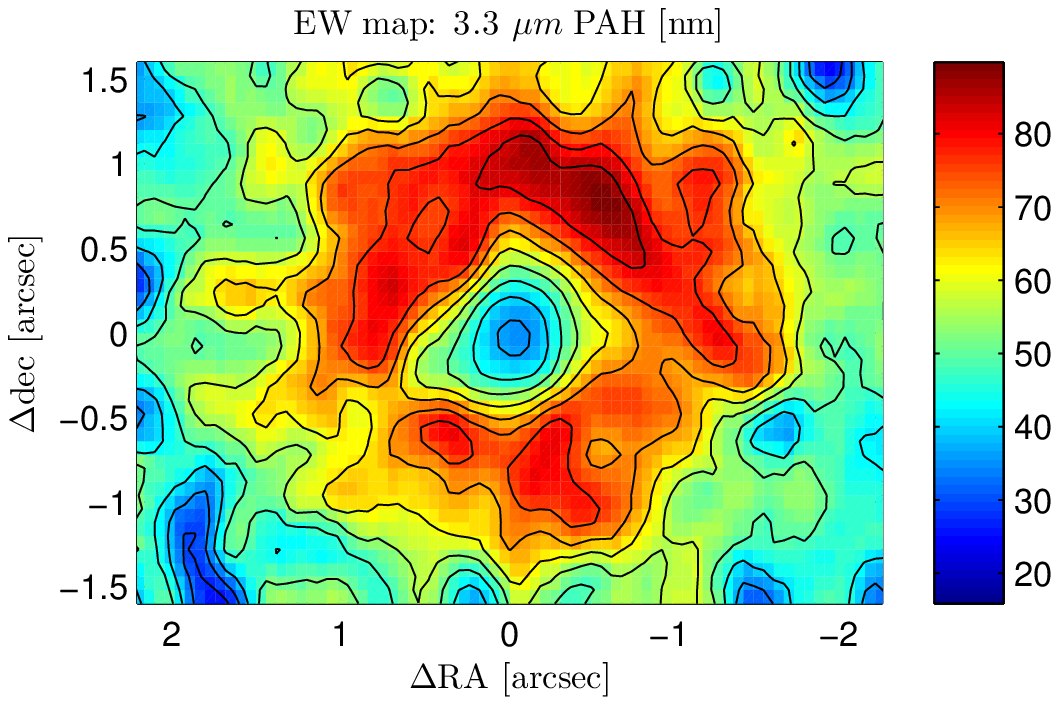}
  \end{minipage}
  \hfill
   \begin{minipage}[t]{150pt}
    \includegraphics[width=8cm,clip=true]{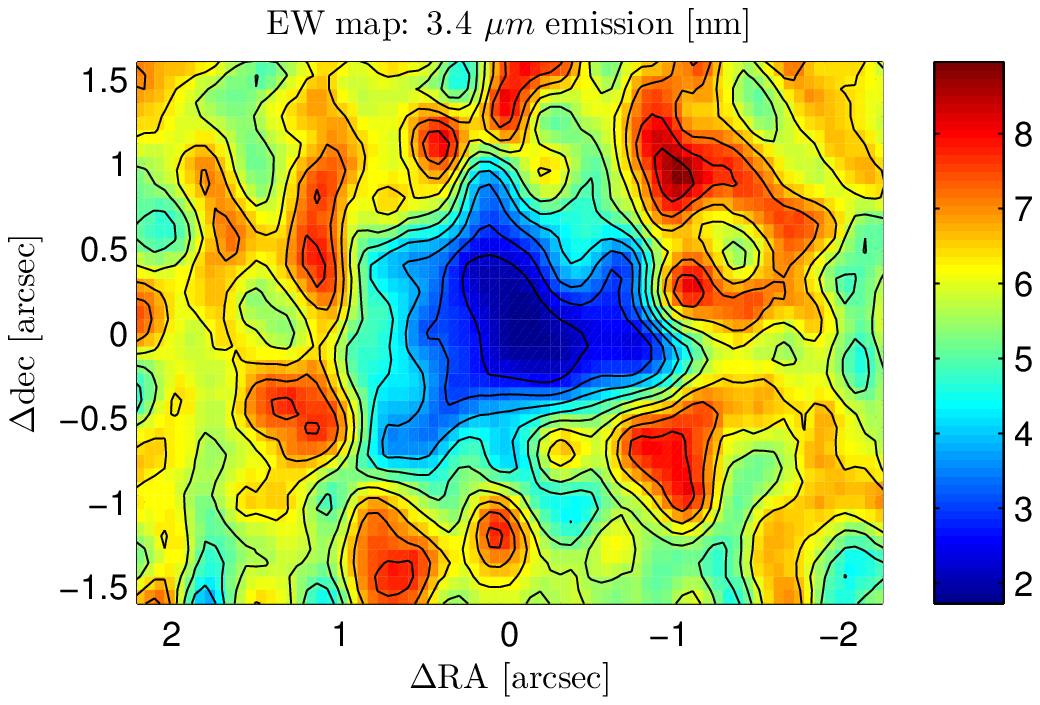}
  \end{minipage}
\caption{\label{PAH12} 
Top: Map of the 3.3 $\mu$m PAH emission flux obtained by summing 
over the feature and subtracting the continuum.   Middle and bottom:
Equivalent width maps of the 3.3 and 3.4 $\mu$m PAH emission features, respectively, 
obtained by summing over the feature, subtracting the continuum, and dividing by the continuum. 
 }
 \end{figure}

\section{Discussion}

\subsection{The nucleus}
\label{nucleus}

The NGC~1614  nucleus has a very high luminosity of $M_H$ and 
$M_K \sim -23$ to $-23.5$ mag within the central 0.7\arcsec\ \citep{Alonso2001}. Consistently, we measure approximately $M_L \sim -24$ from the IRAC 3.6 $\mu$m image for the central source.
The $V-H$ (Vega) colour is approximately 
4.5 mag in the same aperture, using the optical HST images.   The extinction at the
core region is estimated at roughly $A_K \sim 0.3-0.4$ (see e.g.\ \citet{Alonso2001} and 
Section~\ref{agn} below).  We ran SB99 models \citep{Leitherer1999} 
with normal Salpeter IMF between 1-120 $M_{\odot}$ 
and Solar metallicity, and found that with the assumed extinction, the core is consistent with
a 100 to 200 Myr age single stellar population and a mass of $\sim 1 \cdot 10^9 \ M_{\odot}$,
though of course multiple populations are likely in reality, including younger ones.
Nevertheless, the lack of recent strong star formation indicators (radio, Br$\gamma$, Pa$\alpha$) 
in the core consistently suggests it no longer contains ionising O/B stars, and therefore has not
participated in a starburst over the last 50-100\,Myr.  
The dynamical mass of the nucleus in the central 1\arcsec\  to 2\arcsec\  area has been estimated to be
in the range 1 to $2\cdot 10^9 \ M_{\odot}$ \citep{Puxley1999,Alonso2001,Olsson2010}.
Mass determined from and related to recent star-formation is more model dependent, but is also 
close to $1 \cdot 10^9 \ M_{\odot}$ \citep{Kotilainen2001}.  All these values are in
good agreement with our crude estimate above.  The de-convolved size obtained from the  
HST NICMOS images is less than 0.1\arcsec, indicating a radius less than 15\,pc.   
Our $L$-band continuum data suggests that there is an additional, slightly more extended 
component to the nucleus, possibly a result of a radially differentiated stellar population and/or dust distribution -- it is not possible to investigate this further with the present data however.

The mass of the core component places it well above that of typical nuclear star clusters (NC)
in spiral galaxies, which range from $10^6$ to $10^8$ $M_{\odot}$, while the 
size appears to be a factor of $\sim2$ larger than a typical NC \citep{Boeker2002,Rossa2006}.
However, the subtraction of an underlying bulge light is very difficult, and hence 
we would not definitively rule out a 'nuclear cluster' classification, and prefer to talk more 
loosely about a 'core' here.   Within a larger aperture of 100 to 200 pc, the magnitude
and colour of the NGC~1614 core is in fact well within the range measured in the same
aperture for nearby merger remnants from NICMOS imaging \citep{Rossa2007}. The large
mass and blue colour of the NGC~1614 nucleus would be indicative of a well-advanced 
interaction/merger.

\begin{figure*}
\hspace{4mm}
\includegraphics[width=8.45cm,clip=true]{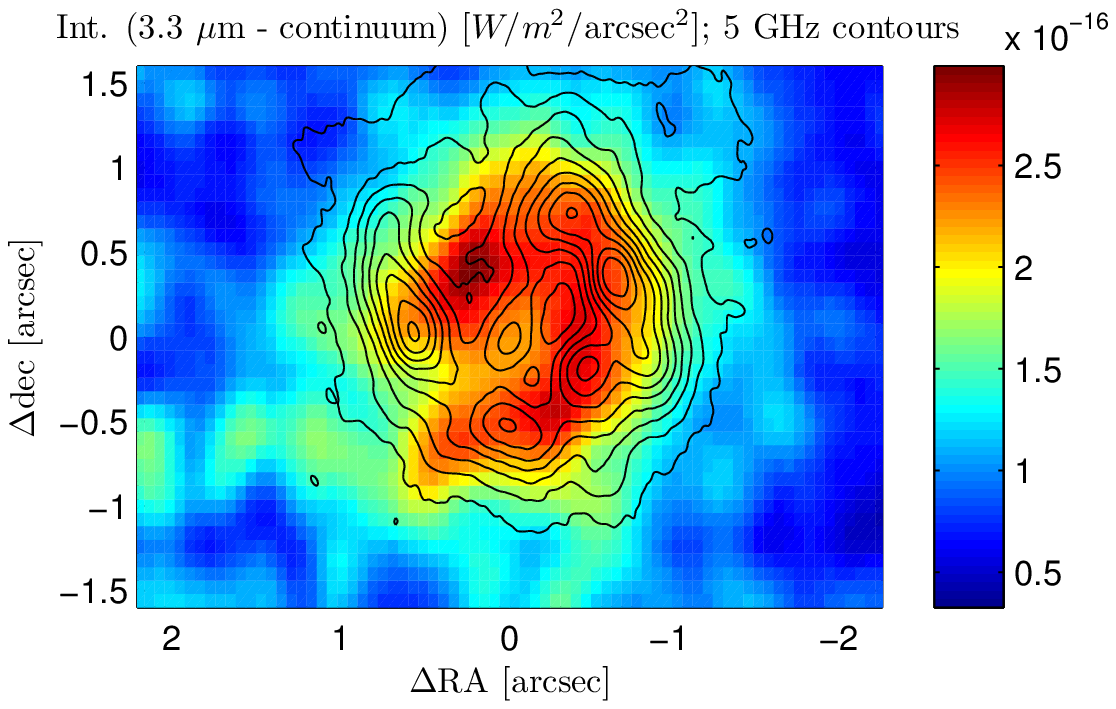}
\hspace{0.5mm}
\includegraphics[width=8.45cm,clip=true]{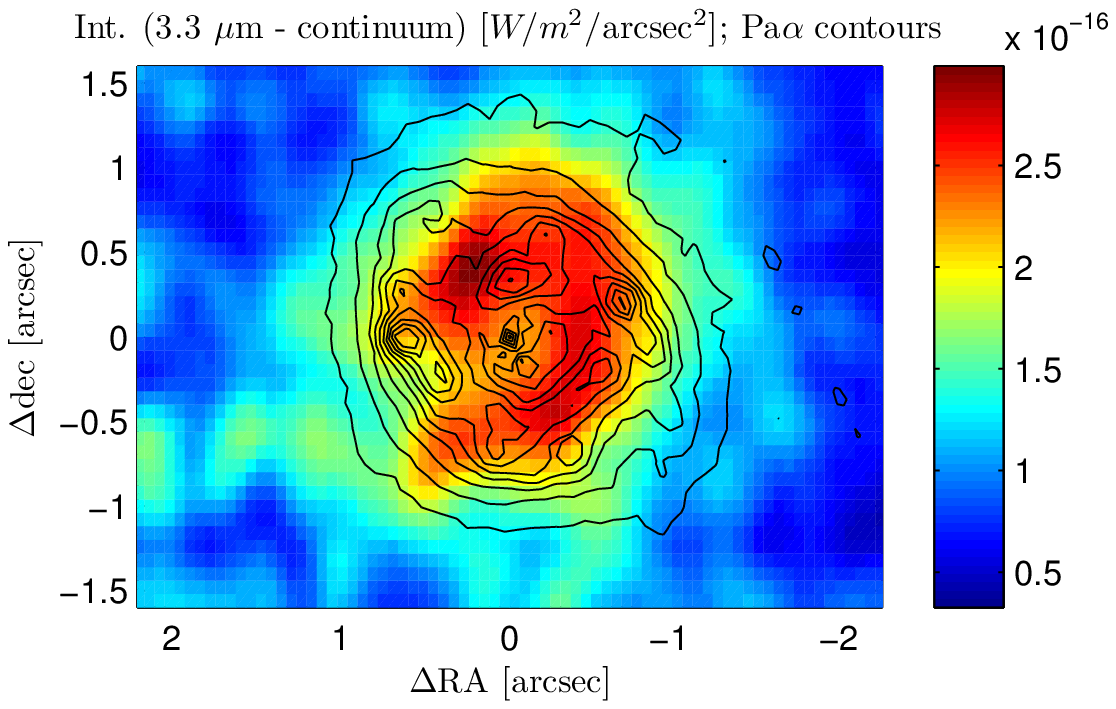}
\caption{\label{sfring2}
The radio and Pa$\alpha$ emission is shown with linearly spaced contours overlaid on the 3.3 $\mu$m PAH map.
}
\end{figure*}

\subsection{Second nucleus?}
\label{secondnucleus}

There have been several reports of a double nucleus in this galaxy, dating
back to \citet{Keto1992}, which was based on a de-convolved 12 $\mu$m
image. \citet{Alonso2001} found a secondary point source (different from the
Keto source) 0.9\arcsec\ NE of the nucleus, and suggest it is a secondary
nucleus, potentially the remnant of a companion galaxy.  

The secondary source of \citet{Alonso2001} is visible in all available HST
images just a little outside of the star forming ring in the area of the 
PAH emission found in this work. It is unresolved, and much bluer than the infrared nucleus itself. At
1.6 $\mu$m, it has become faint.  Its $V-H$ colour is $\sim 0.6$, some 4 mag bluer than that
of the main nucleus, while still being an intrinsically bright source of $M_H\sim -18$. 
The newer archival ACS data show $B-I$ $\sim 0.5$.  This secondary source is within the UIST 
field of view but is not visible in the data,  confirming its blue colour and indicating low extinction.  
Using the same SB99 single stellar population model as in the previous section,  
we constrain the age of the blue point source to be less than 10 Myr in the absence of any extinction.  
With even small dust extinction its age would only be several million years. 
Given the large number of star clusters in this system elsewhere, the most likely explanation 
for this source is a young star cluster, in fact the brightest one in NGC~1614, and probably lying 
in the foreground in relation to most of the nuclear dust and gas.  
 The MIR image from \citet{Keto1992} showed two components 0.6\arcsec\
apart.  Given the spatial resolution and the used 12-$\mu$m filter which combines PAH
and continuum emission, this image may show the central (continuum) core with an emission 
peak in the PAH ring (see Fig.~\ref{PAH12}).
Our UIST data and new archival HST/ACS data therefore do not provide evidence for a 
double nucleus in the central region of this galaxy.

\begin{figure}
\includegraphics[width=8cm,clip=true]{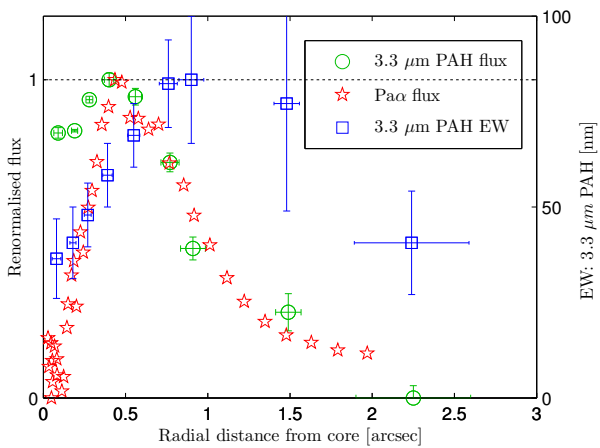}
\caption{\label{radplot} The EW of the 3.3 PAH, the PAH feature flux, and Pa$\alpha$ flux
shown as a function of radius. Each profile is normalised to unity.  However, the actual values
of the 3.3 PAH EW can be read from the right-side axis.}
\end{figure}

\subsection{Companion galaxy}
\label{companion}

Instead of the double nucleus scenario, we propose rather 
that the companion galaxy (remnant) involved in the 
NGC~1614 system interaction lies within the conspicuous linear tail extending 
more than 20 kpc South from the central regions (Fig.~\ref{hstpics}, upper right panel).
In this picture, the whole linear tidal tail would be the edge-on remains of a precursor disk galaxy.

The archival HST images show a smooth elongated region resembling a
dwarf galaxy, or a stretched-out bulge, with a bright stellar cluster at its centre
some some 10\arcsec\ (3kpc) from the galactic nucleus along the linear tail.
This structure clearly stands out from its surroundings and is perpendicular to 
the underlying major spiral arms, and \citet{Neff1990} note the discrepancy between its 
kinematic structure and the rest of the system.  The feature is outside the FOV of the NIR images, 
but we determined from the archival ACS images that 
the cluster has a magnitude of $M_I \sim -14$ mag, 
and a $B-I$ colour of approximately 1.8.  An elliptical aperture of $2 \times 4\arcsec$ 
($0.6 \times 1.2$~kpc) extracted around the feature, against the backdrop of the tidal tail, 
gives $M_I \sim -17$ mag,  while the colour of the feature is similar to that of the cluster.
The colours of both the cluster and the extended feature as a whole suggest evolved 
stellar populations.  The cluster is of typical spiral NC size and luminosity, strengthening the idea
that this knot would in reality be the remnant of a spiral galaxy responsible for the tidal 
features of NGC~1614.  It is impossible to estimate the original mass of the secondary galaxy 
from the available data, but its extracted 'bulge' flux is only a few per cent of the primary nucleus.
Because of the presence of a young starburst in the primary nucleus, the $I$-band light can not be 
trusted to give any reliable indication of the stellar mass ratio, however.

A plausible scenario according to dynamical simulations including stellar, gas and dark matter
dynamics and star formation and various feedback processes (Peter H. Johansson, 
private communication; \citet{Peter2009a}) 
would be that of a merger with a mass-ratio
in the range $\sim$ 5:1 to 3:1, where the 
smaller galaxy, seen edge on, has initially passed the primary component producing the largest scale
tidal tails.   This interpretation is supported by the fact that we see strong tidal features, and the 
companion galaxy appears to have a significantly  lower mass:
the observed phase would be $\sim 50$~Myr after the {\em second} passage, 
which has triggered the strongest nuclear starburst, and the minor component has lost most of its 
stellar mass, and its rotationally supported former disk stars have formed the linear tail. 
The bulge region of the companion with its deeper potential well is more intact and is falling towards
 the primary nucleus for the third time to merge in another 100 Myr or so. 
Exact timescales are dependent on geometry and gas fractions.  A detailed specific 
simulation is  beyond the scope of this paper, 
and in addition observational velocity data would be required
to constrain  the details in the numerical simulation. An interesting detail would be to find out
what size and what kind of a companion could simultaneously trigger LIRG activity and tidal tails at 
this level and leave so little evidence of itself.  Simulations do not predict significant SFR elevation
in mergers beyond $\sim$5:1 mass ratios \cite[e.g.][]{Cox2008,Johansson2009b}. 
And intriguing would be an investigation of the supermassive black hole (SMBH) which
would be expected to be growing already in the centre with more material falling there as the  
the interaction progresses.

\subsection{The missing AGN}
\label{agn}

AGNs give rise to a strong, inverted continuum in the L band. The
$L$-band {\it continuum} emission from an AGN is a factor of 100 stronger than
that of a starburst \citep{Risaliti2006}: an AGN dominates the $L$-band
continuum, even when its contribution to the bolometric luminosity is
relatively minor \citep{Imanishi2000,Sani2008}.  Starbursts in turn 
yield strong 3.3$\mu$m PAH emission (C-H stretching vibration), related to UV radiation 
and shocks, while AGNs destroy PAH molecules
through their X-ray emission \citep{Voit1992} unless shielded by significant
dust and molecular clouds. Thus, the PAH equivalent width
distinguishes starburst regions from AGN dominated
regions \citep[e.g.][]{Imanishi2006} even with a strong 
AGN dominance in the continuum.  The cool dense gas
around an AGN can show a 3.4 $\mu$m absorption feature, due to aliphatic
hydrocarbons, and if shielded molecular clouds are present, also a 3.1 $\mu$m
ice band. No other single spectral region combines such diverse tracers, and the
$L$-band IFU data used here are therefore ideal for SF vs. AGN studies.

The debate about the existence of an AGN in NGC~1614 has remained inconclusive
in the literature.   \citet{Risaliti2000} detect 
an X-ray source in the centre of NGC~1614, the spectrum of which they interpret as 
being obscured by Compton-thick gas.  \citet{Olsson2010} point out that this detection
could also be explained by low-mass X-ray binaries; their own radio data meanwhile
show a weak central radio source which could in principle be due to AGN activity.

The lack of any significant AGN detection can in principle be either because of very low 
luminosity or because of extreme obscuration.  \citet{Risaliti2006}, however, argue that
even a 1 percent bolometric contribution from an unobscured AGN would be clear in the 
$L$-band continuum.  
$L$-band spectra can contain three diagnostics of AGN activity: weak 3.3 $\mu$m
emission, 3.4 $\mu$m absorption, and/or a rising continuum. 
None of these are found in our data of NGC~1614.  The PAH band is strong,
albeit weaker towards the central infrared core.  The continuum is flat, and
the 3.4-$\mu$m band is in emission in the star-forming ring, and absent
elsewhere (see Section~\ref{ifuspectra}).

\begin{figure}
\includegraphics[width=8.0cm,clip=true]{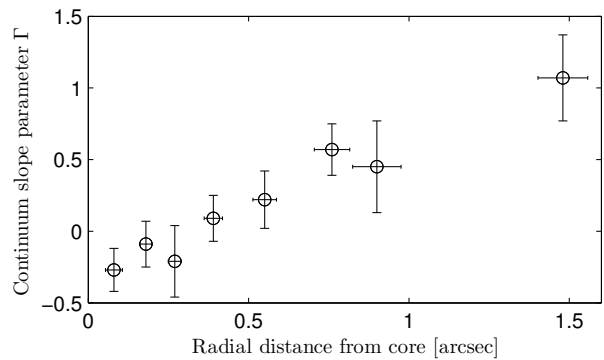}
\caption{\label{radplots2}  Right:  The $L$-band continuum shape parameter $\Gamma$
as a function of radius.}
\end{figure}

More quantitatively, when the $L$-band continuum is parameterised with its spectral index
$f_{\lambda} \sim \lambda^{\Gamma}$, \citep[see][]{Risaliti2006},  
obscured AGN lie at $\Gamma > 1.5$, i.e.  the continuum is significantly reddened.
Starbursts and unobscured AGN both have lower $\Gamma$, typically in 
between -1 and 1.  Furthermore, starbursts and unobscured AGN can be 
differentiated by high vs. low  EW$_{3.3}$ values, respectively, where the dividing 
line is close to EW$_{3.3} \sim 75$ nm. 
The EW of the 3.3 PAH feature in our data over the whole fov is 
EW$_{3.3} \sim 60\pm12$ nm, 
and the shape of the continuum $\Gamma = 0.31\pm0.19$.
The $\Gamma$ value places NGC~1614  
far away from the obscured AGN, but the 3.3 $\mu$m EW is 
close to the dividing line of SBs and 
unobscured AGN.  \citet{Risaliti2010} note that many LINERs are found in this
region of the diagnostic.  AKARI spectroscopy \citep{Imanishi2010} gives a somewhat 
larger EW$_{3.3} \sim 120$ in a much larger aperture, while the continuum 
slope is the same as in our widest aperture.   The PAH EW as a function of radial
distance from the core is shown in Fig.~\ref{radplot} and the continuum parameter
in Fig.~\ref{radplots2}:  it is the nucleus which gives
AGN-like EW values, counteracting the more SB-like values from the PAH ring. 
In fact, regarding only the $<0.3$\arcsec\ aperture in the centre, these conventional
$L$-band parameters would place the NGC~1614 in the unobscured 
AGN regime.  

Extreme obscuration does not appear to be the cause of lack of an 
AGN signature in our $L$-band data, consistent also with the non-detection of
any 3.4-$\mu$m absorption.  Additionally, this is consistent with extinction estimates for 
NGC~1614 in the literature, where derivations using NIR lines  
\citep{Puxley1999, Kotilainen2001, Alonso2001}  estimate $A_K < 1$ for the whole 
central region, and in fact closer to $A_K \sim 0.4$ right at the centre.
The Spitzer IRS spectrum potentially probes deeper into the NGC~1614 core,
albeit with a much lower spatial resolution. Firstly, the strength of the silicate
absorption feature can be parameterised as 
$S_{\rm Sil} = \ln (F_{9.7,{\rm cont}} / F_{9.7,{\rm abs}} )$ and we find a value of  
$\sim -0.4$, which together with the 
continuum levels at 14 and 30 $\mu$m translates to a likely nuclear extinction
of $A_V \sim 15-20$ \citep[][]{Levenson2007}, i.e.\ $A_K \sim 2$ or 
$A_L \sim 1$, which is quite modest compared to most ULIRG nuclei.
Second, SB vs. AGN diagnostics such as [\mbox{N\,{\sc V}}]/[\mbox{N\,{\sc II}}] 
($\sim 0.04$), [\mbox{N\,{\sc III}}]/[\mbox{N\,{\sc II}}] 
($\sim 0.2$) and EW$_{6.2}$ ($\sim 0.5$ $\mu$m) relative to the strength of the silicate 
feature $S_{\rm Sil}$ $(\sim -0.4)$ all indicate pure
starburst values, similar to e.g.\ M82 \citep[e.g.][]{Farrah2007,Spoon2007}.  

If indeed there was Compton-thick gas of column density $\log N_H > 24$ obscuring an AGN, it
would indicate, using standard gas-to-dust conversions, nuclear extinctions of 
$A_K > 40$.  PAHs in the NGC~1614 nuclear regions can co-exist there
with an AGN, if e.g.\ simultaneously mixed and shielded with a thick dusty (unresolved) torus
\citep{Watabe2008}.  However, in this case we 
should see the $\Gamma$-parameter rise, or at least flatten, toward the centre since we can isolate
a $\sim30$ pc region in the centre.   There is not even a hint of this (Fig.~\ref{radplots2}), nor
any signs of extreme dust extinction within the resolvable area, and we thus conclude that an 
obscured AGN is definitely ruled out in NGC~1614.

The final option, in principle, would then be that there is an unobscured AGN 
in the centre.  It would have to be very weak, however,  given the
non-detection of such in the X-rays.  Furthermore, while the EW of the PAH emission 
is weaker in the centre, some PAH flux is nevertheless detected throughout the 
nuclear region (Fig.~\ref{PAH12}).  An unobscured AGN would be expected  to
destroy the PAH carriers at least within the central kpc.  Instead, it is much more
plausible that the weakened EW$_{3.3}$ is due to the central starburst (see
Section~\ref{ring2}).

From the point of view of the typical ULIRG to QSO to elliptical evolutionary sequence, 
it is somewhat surprising that a strong LIRG at an advanced merger stage does not 
show any signs of an AGN. 
Gas rich mergers such as NGC~1614 are expected to produce both high SFRs and
the most rapidly growing SMBHs 
(Johansson et al.\ 2009a, Johansson et al. 2009b).
As discussed in the previous section, we appear to be dealing with a case of a well advanced
major merger in the 3:1 to 5:1 mass ratio range.  
Another interesting question for a future simulation would thus be an 
investigation into why a SMBH does not activate, while simultaneously producing a nuclear 
starburst and tidal features typical of major mergers.

\subsection{The starburst ring in NGC 1614}

\subsubsection{Ring stratification}
\label{ring1}

Circumnuclear star-forming rings (or tightly wound spiral arms) are a common
feature of barred spiral galaxies \citep[e.g.][]{Boeker2008}. They are found in
similar proportions in both starbursts and Seyfert galaxies with a
star-formation contribution \citep{Kotilainen2000}. 
The rings appear
to be related to the presence of dynamical resonances with the bars, with gas
accumulating at the inner Lindblad resonance, typically at about 1 kpc radius
from the centre, and experiencing a starburst through collision of molecular
clouds or through gravitational collapse. Examples of recent studies of such
rings are found in e.g.\ \citet{Beck2010, Reunanen2010,Hsieh2011}. On the other hand,
SF rings might be a result of a ``classic" nuclear starburst propagating outward, where the
nuclear regions have been left devoid of most gas. These strong nuclear
starbursts are expected to result in interactions and mergers, even minor mergers
 \citep[e.g.][]{Mihos1994}. 

Though there are many long-slit spectroscopic studies of galactic nuclei in the 
3.3 $\mu$m PAH band \citep[recently e.g.][]{Imanishi2006,Risaliti2006,Oi2010}, 
studies with full spatial information of the 3.3 $\mu$m PAH emission in the central regions 
of galaxies are not common. \citet{Cutri1984} and \citet{Mazzarella1994} find a hint
of a ring-like structure around the Sy1 nucleus of NGC~7469 in narrow-band 3.3 $\mu$m 
imaging, while \citet{Tacconi2005} study two nearby starbursts NGC~253 and
NGC~1808 and note the PAH feature to peak around the strongest SF regions. 
The nuclear 3.3 $\mu$m PAH ring found and characterised in this work with
IFU data in NGC~1614 is by far the highest signal-to-noise detection
of a ring feature at this wavelength around a major galaxy nucleus.  
Below we discuss this PAH ring in NGC~1614 to further characterise the ring structure
and SF in the galaxy's nuclear region.

The circumnuclear star-forming ring in NGC~1614 was discovered by \citet{Heisler1999}, 
\citet{Alonso2001} and \citet{Kotilainen2001}, in Br$\gamma$, Pa$\alpha$, 
and near-infrared extinction, respectively. \citet{Olsson2010} found it in radio continuum.
The presence of radio continuum and hydrogen line emission shows that the gas is
ionised, and forms one, or a clustering of, \mbox{H\,{\sc ii}} regions, confirming 
ongoing or recent star formation.  
 The ring in NGC~1614 is smaller than usual, however, at 150\,pc radius.  There is
also no clear evidence for a bar \cite[though see][]{Olsson2010}. 
Instead, the ring seems more similar to those found in
ULIRGs \citep{Downes1998} which range in radius from
300 to 900\,pc, and exist in apparently chaotic surroundings.

The Br$\gamma$ emission in the nuclear regions of NGC~1614 comes from the 
inner edge of the extinction ring \citep{Kotilainen2001}, and \citet{Alonso2001}
show that the Pa$\alpha$ emission peaks slightly inside of the long-slit detected 
H$_2$ warm gas distribution.  A similar case where star formation is happening on the 
inside of a dense molecular ring of gas is found by \citet{Combes2004} in NGC~7217.

In Section~\ref{radialdeps} and Fig.~\ref{radplot} 
we showed that the radio and Pa$\alpha$ emission 
peak just inward of the PAH emission found in this work leading to the picture
of a stratified structure of gas.  All this
strongly suggests that the star formation in the nuclear regions of NGC~1614 
occurs at the inner edge of the gas distribution, i.e.\ the majority of the dense gas trapping 
the photo-dissociation region is located outward.

\subsubsection{Propagation of star-formation in the ring}
\label{ring2}

\citet{Olsson2010} present a model where the SF  ring is a dynamical resonance 
region - the SF is tracing the leading edge of a bar, and gas piles up at the ring location. 
Another inner bar, suggested by the radio data, would further funnel gas into the 
centre as well.  This scenario might be expected in a traditional bar-driven spiral nucleus, 
though minor mergers may also trigger inner rings \citep{ElicheMoral2011}.   
In contrast, \citet{Alonso2001} suggested a nuclear starburst scenario, which is often associated
with major mergers and interactions.  In this case the SF would propagate from 
the centre outward into the molecular gas seen as the $J-H$ extinction shadow and  
H$_2$ distribution \citep[also][]{Kotilainen2001}.  The derived ages from NIR 
spectroscopy of the starburst support this; they find a best fit with a double-peak 
starburst with the older ($>10$ Myr) starburst in the centre and the younger (5-8 Myr) starburst
possibly in the ring.  \citet{Olsson2010} agree with these values based on radio spectral indices;
however, they propose, based on their CO-data, that the extinction shadow is not a ring at all, 
but due to an off-centre foreground dust lane, North-West of the nucleus. 
Based on the velocity structure of their CO-data they further speculate on the existence of a 
nuclear (decoupled) North-South bar in the nucleus, which would be responsible for funnelling
gas to the central starburst.  
 The PAH ring discovered in this work, however, is very consistent with the positioning of 
 the previous  NIR detected SF ring and its stratified structure. The differences between the CO 
 and NIR structures could rather indicate the different distributions of the 
 warm molecular gas and the cool gas.  Also, neither our UIST data nor the ACS 
 $B$ and $I$-band images hint at an inner bar, though the complicated dust structures in the
 nucleus (see Fig.~\ref{hstpics}) would in any event make interpretations difficult. 

\citet{Tacconi2005} show evidence that the 3.3 $\mu$m PAH feature-to-continuum ratio 
(essentially EW) is high {\em around} strong SF such as super star clusters and nuclei, but
is low specifically {\em at} those strong SF locations.   
This is totally consistent with what we see here: strong EW of PAH in a 
ring just {\em outside} of the SF ring.  The PAH EW could be weak either
because of destruction of the carriers,  or because of dilution of the feature by a strong 
continuum. The actual weakening of the PAH signal, in contrast to dilution by continuum light, 
could be through either photoionisation or photodissociation of the PAH molecules in a sufficiently
intense radiation field, or through mechanical energy input via stellar winds and 
supernovae \citep[see e.g.][and references therein]{Tacconi2005,Mason2007,Lebouteiller2011}

Meanwhile,  \citet{Diazsantos2008} find that the youngest \mbox{H\,{\sc ii}} 
regions in their sample of several LIRGs show the lowest 8 $\mu$m and 11.3 $\mu$m PAH 
contributions; specifically, they find that the PAH-to-P$\alpha$ ratios grow with the age of the 
PAH emitting region. In fact, their sample includes NGC~1614, and the P$\alpha$ ring is seen
to correspond to an annular 'hole' in a 8$\mu$m / Pa$\alpha$ map.  Their data do not go
far enough in radial distance to detect the PAH ring we see in our data.  
Figure~\ref{radplot} shows that the PAH flux to Pa$\alpha$ ratio (ratio of green circles to red 
stars in the plot) has a wide minimum at the SF ring radius from about 0.4 to 0.8\arcsec.
Interestingly, the ratio rises very rapidly at the inner edge of the SF ring towards the 
core, consistent with older SF.  The PAH EW drops there as well, while on the outward
side of the ring the EW keeps rising, reaching its peak only at the outer edge of the SF ring.
How to reconcile these observations?   Firstly, \citet{Diazsantos2010}
suggest a correlation with the mass of the starburst region in a sense that lower participating 
masses (i.e. the outer, plausibly less dense regions of the molecular gas ring) at a given age
 increase the measured strength of the PAH.  More significantly,  \citet{Beirao2009} find that 
 H$_2$ emission 
in Arp~143 traces a shock wave region outside of the nucleus, with the {\em very}  youngest (2-4 Myr) 
SF knots corresponding to significant 8 $\mu$m PAH emission, while the older (7-8 Myr) SF 
knots are more devoid of PAH.   Could these very young regions correspond to our PAH EW
ring just outward of the SF ring? It is 
plausible, and could in fact be checked with an age and mass modelling study of the numerous 
SSCs found in the nuclear regions in the archival HST data; such a study is, however, beyond the
scope of this work.  
Nevertheless, the location of the PAH ring found in our UKIRT IFU data appears to fit 
very well the scenario of an outward propagating starburst. 

Returning to the MIR emission in the very core of NGC~1614, all indicators show that the starburst
 population there is older (Section~\ref{nucleus}).  
The nucleus appears to be surrounded by a hole, or at least a depression, in the ISM, as shown 
by the decreasing extinction towards the inner regions \citep{Kotilainen2001} and the weakening of
the PAH emission inside of the star formation ring. A recent starburst in the
nucleus could have evacuated the inner regions through mechanical energy due to stellar winds 
and supernova eruptions.  This event has likely happened 20 to 100 Myr  ago, given the
lack of current SF in the core, as well as the best-fitting ages of the modelled colours 
and emission lines in the core.  There is, nevertheless, PAH emission even over the core region, 
which could indicate that the PAHs there trace at least partially an older ($\sim$100 Myr timescale) 
star formation of  B-type stars \citep{Peeters2004}.  
We note also that the 12.5 and 18 $\mu$m warm dust emission in the NGC 1614 core region is spatially extended in a $\sim2$ \arcsec\ diameter region \citep{Soifer2001, Imanishi2011}, i.e. similar in extent than the 3.3 $\mu$m PAH flux seen in Figs.~\ref{sfring2} and~\ref{radplot}.

A consistent picture thus emerges: viz., one where the nucleus has experienced
a starburst more than 10 Myr ago, perhaps several such episodes,
resulting in a mixed stellar population, and a depletion of PAHs and ISM in the central 100 pc 
radius.  The timing of the second passage of the secondary companion $\sim50$ Myr ago 
discussed in Section~\ref{companion} is consistent with this timescale.  
The starburst is propagating outward, and the strongest current SF is happening in the 
Br$\gamma$, Pa$\alpha$ and radio continuum detected star-forming ring between 120 to
240 pc, where the PAH carriers are also being destroyed.  
The detection of a high-EW PAH region just outside of this SF ring at 200 to
550 pc radius, overlapping with a distribution of H$_2$ gas, is consistent with the starburst being 
currently expanding into this ring of the densest molecular gas.  We speculate that the PAH ring
thus traces the region in the system with the most recent SF, possibly a shock front.  Major 
galactic scale winds would be expected in a case like this.  Indeed the detection of significant 
outflowing cool gas, seen as a 150 km/s blue-shifted NaD component from the nucleus
of NGC~1614 \citep{Schwartz2004} supports the scenario -- in fact, a red-shifted in-falling
 component is detected as well, and the decoupling of cool and warm gas flows, highlighting the
fact that NGC~1614 is a complex interaction/merger case.

Finally, we note that in NGC~1614, PAH emission in the 3.3 $\mu$m band  
is detected in a 'wide' area in and around the nucleus.
PAH emission is found at the very core, in the strongest SF ring, 
as well as outside it. The PAH emission does not trace only the most 
massive SF, but rather a variety of levels of SF,  the EW of the PAH outside of the
SF ring possibly detecting SF levels not reached by other SF indicators.
The spread out nature of the PAH emission within different kinds of 
nuclear components indicates that the PAH carriers are quite mixed within the obscuring 
material, rather than tracing for example only the densest obscured SF regions -- this result 
was recently found by \citet{Zakamska2010} to apply for ULIRGs in general.

\subsection{Future of MIR IFU}

IFU studies in the mid-IR have great relevance for starburst and AGN
studies, where various emission and absorption features can be used effectively 
to differentiate between different excitation and heating mechanisms and physical sources
of energy.  Lack of spatial information, however, means that one is often forced to analyse and 
attempt to decompose integrated signals.  This work shows the 
importance of simultaneously obtaining spectral and spatial information to disentangle 
physical processes in the cores of galaxies from those happening in the close vicinity.
Sadly, there currently exists no instrument to perform spectral and spatial 
$L$-band studies; UIST was the only available MIR IFU for 
 the astronomical community.  However, with the advent of ELTs and JWST 
 and associated instruments \citep[e.g.][]{Closs2008,Kendrew2010}, 
 these observations will once again become possible.  
 And while the current ground-based telescopes can achieve resolutions of
 $\sim 0.2$\arcsec\ in the MIR, or  $\sim20$ pc with a few of the closest AGN and starbursts, the 
 ELTs will extend the spatial resolution to well below 10 pc for hundreds of starbursts and AGN.
 Thus, in the future, one can use these same MIR IFU methods to examine and characterise 
 SF in the putative molecular tori around the central black holes.  For example, \citet{Oi2010} and 
 \citet{Watabe2008} advocate that there should be SF in the outer parts of these tori,
 which themselves shield the dust and PAH carriers from the central engines. 
 Observations to test these suggestions directly are still unreachable with current resolutions.

\section{Conclusions}

We have presented new UKIRT/UIST $L$-band integral field unit imaging spectroscopy of the central 
kpc region of the luminous IR galaxy NGC~1614.  In particular, we studied the spectral and 
spatial distribution of the 3.3 $\mu$m PAH feature and the related continuum.

1)  The main observational result shows that the equivalent width map of the 3.3 $\mu$m PAH 
feature forms a ring around the
central continuum source.  This ring peaks at 250 pc radial distance from the core, and 
extends from around 200 to 500 pc.  The 3.3 $\mu$m PAH flux, on the other hand, is somewhat
more centrally distributed, though it also has a depression at the core.

The strongest PAH emission, as traced by the EW,  peaks outside the previously detected 
star formation (radio, Br$\gamma, $Pa$\alpha$) ring, 
while the actual PAH flux is more or less coincident with the SF ring, or slightly inward of it.
Unlike the other SF indicators, PAH flux is also detected all the way to the centre in our 
$\sim$0.3\arcsec\ resolution data. The PAH emission appears to trace a variety of
levels of SF in the nuclear regions of NGC~1614.

2) The second main observational result is that we rule out the existence of an obscured AGN in
NGC~1614 using spatially resolved $L$-band SF vs. AGN diagnostics.  
We also find it extremely unlikely that an unobscured AGN would be present, leaving NGC~1614 
a pure starburst galaxy. 

3)  Using archival HST/ACS data we argue that the likely remnant of a companion galaxy 
in the NGC~1614 interaction is not found in the primary nuclear region as previously suggested, 
but rather approximately 3 kpc from the primary core along the South-West linear tidal tail.
Comparing with numerical simulations we find that the observed system is consistent with 
a relatively major unequal-mass merger with a mass-ratio in the range of $\sim$5:1-3:1.  
In this scenario NGC~1614 is observed $\sim50$ Myr 
after the second passage and the edge-on secondary is responsible 
for the long linear tail of the system and has lost most of its mass.

4)  The characteristics of the starburst in the centre (no current SF in the core, strong SF in a
ring) and the location and equivalent width characteristics of the PAH we detect, as well
as the timing of the likeliest interaction scenario, all fit well a picture of an outward propagating
starburst, triggered most probably by the passage of the companion galaxy.   The outstanding
question is why there are not yet any signs of an AGN in this advanced merger, as might be
expected in typical gas-rich spiral to (U)LIRG to QSO/elliptical evolutionary scenario.

\section*{Acknowledgements}

We gratefully acknowledge support and hospitality from the Astronomy Division at the University of 
Helsinki and the Finnish Centre of Astronomy with ESO at Tuorla observatory.   
We thank Peter H. Johansson for
helpful comments and discussion regarding dynamical galaxy simulations, and the referee for
useful comments.  PV acknowledges financial support from the National Research Foundation.
JK acknowledges financial support from the Academy of Finland, 
projects 8107775 and 2600021611. The data reported here
 were obtained as part of the UKIRT Service Programme, Chris Davis who 
carried out the observations and provided subsequent support is especially thanked. 
The United Kingdom Infrared Telescope is operated by the Joint Astronomy
Centre on behalf of the Science and Technology Facilities Council of the
U.K.

\bibliographystyle{mn2e}

\bibliography{n1614_ref}

\begin{thebibliography}{}

\bibitem[\protect\citeauthoryear{{Alonso-Herrero}, {Engelbracht}, {Rieke},
  {Rieke} \& {Quillen}}{{Alonso-Herrero} et~al.}{2001}]{Alonso2001}
{Alonso-Herrero} A.,  {Engelbracht} C.~W.,  {Rieke} M.~J.,  {Rieke} G.~H.,
  {Quillen} A.~C.,  2001, \apj, 546, 952

\bibitem[\protect\citeauthoryear{{Alonso-Herrero}, {Rieke}, {Rieke}, {Colina},
  {P{\'e}rez-Gonz{\'a}lez} \& {Ryder}}{{Alonso-Herrero}
  et~al.}{2006}]{Alonso2006}
{Alonso-Herrero} A.,  {Rieke} G.~H.,  {Rieke} M.~J.,  {Colina} L.,
  {P{\'e}rez-Gonz{\'a}lez} P.~G.,    {Ryder} S.~D.,  2006, \apj, 650, 835

\bibitem[\protect\citeauthoryear{{Arribas}, {Colina}, {Monreal-Ibero},
  {Alfonso}, {Garc{\'{\i}}a-Mar{\'{\i}}n} \& {Alonso-Herrero}}{{Arribas}
  et~al.}{2008}]{Arribas2008}
{Arribas} S.,  {Colina} L.,  {Monreal-Ibero} A.,  {Alfonso} J.,
  {Garc{\'{\i}}a-Mar{\'{\i}}n} M.,    {Alonso-Herrero} A.,  2008, \aap, 479,
  687

\bibitem[\protect\citeauthoryear{{Barnes} \& {Hernquist}}{{Barnes} \&
  {Hernquist}}{1992}]{Barnes1992}
{Barnes} J.~E.,  {Hernquist} L.,  1992, \araa, 30, 705

\bibitem[\protect\citeauthoryear{{Beck}, {Lacy} \& {Turner}}{{Beck}
  et~al.}{2010}]{Beck2010}
{Beck} S.~C.,  {Lacy} J.~H.,    {Turner} J.~L.,  2010, \apj, 722, 1175

\bibitem[\protect\citeauthoryear{{Beir{\~a}o}, {Appleton}, {Brandl}, {Seibert},
  {Jarrett} \& {Houck}}{{Beir{\~a}o} et~al.}{2009}]{Beirao2009}
{Beir{\~a}o} P.,  {Appleton} P.~N.,  {Brandl} B.~R.,  {Seibert} M.,  {Jarrett}
  T.,    {Houck} J.~R.,  2009, \apj, 693, 1650

\bibitem[\protect\citeauthoryear{{Bernard-Salas}, {Spoon}, {Charmandaris},
  {Lebouteiller}, {Farrah}, {Devost}, {Brandl}, {Wu}, {Armus}, {Hao}, {Sloan},
  {Weedman} \& {Houck}}{{Bernard-Salas} et~al.}{2009}]{Bernard2009}
{Bernard-Salas} J.,  {Spoon} H.~W.~W.,  {Charmandaris} V.,  {Lebouteiller} V.,
  {Farrah} D.,  {Devost} D.,  {Brandl} B.~R.,  {Wu} Y.,  {Armus} L.,  {Hao} L.,
   {Sloan} G.~C.,  {Weedman} D.,    {Houck} J.~R.,  2009, \apjs, 184, 230

\bibitem[\protect\citeauthoryear{{B{\"o}ker}, {Falc{\'o}n-Barroso},
  {Schinnerer}, {Knapen} \& {Ryder}}{{B{\"o}ker} et~al.}{2008}]{Boeker2008}
{B{\"o}ker} T.,  {Falc{\'o}n-Barroso} J.,  {Schinnerer} E.,  {Knapen} J.~H.,
  {Ryder} S.,  2008, \aj, 135, 479

\bibitem[\protect\citeauthoryear{{B{\"o}ker}, {Laine}, {van der Marel},
  {Sarzi}, {Rix}, {Ho} \& {Shields}}{{B{\"o}ker} et~al.}{2002}]{Boeker2002}
{B{\"o}ker} T.,  {Laine} S.,  {van der Marel} R.~P.,  {Sarzi} M.,  {Rix} H.-W.,
   {Ho} L.~C.,    {Shields} J.~C.,  2002, \aj, 123, 1389

\bibitem[\protect\citeauthoryear{{Bushouse}}{{Bushouse}}{1986}]{Bushouse1986}
{Bushouse} H.~A.,  1986, \aj, 91, 255

\bibitem[\protect\citeauthoryear{{Cavanagh}, {Jenness}, {Economou} \&
  {Currie}}{{Cavanagh} et~al.}{2008}]{Cavanagh2008}
{Cavanagh} B.,  {Jenness} T.,  {Economou} F.,    {Currie} M.~J.,  2008,
  Astronomische Nachrichten, 329, 295

\bibitem[\protect\citeauthoryear{{Closs}, {Ferruit}, {Lobb}, {Preuss}, {Rolt}
  \& {Talbot}}{{Closs} et~al.}{2008}]{Closs2008}
{Closs} M.~F.,  {Ferruit} P.,  {Lobb} D.~R.,  {Preuss} W.~R.,  {Rolt} S.,
  {Talbot} R.~G.,  2008, in Society of Photo-Optical Instrumentation Engineers
  (SPIE) Conference Series Vol.~7010 of Society of Photo-Optical
  Instrumentation Engineers (SPIE) Conference Series, {The Integral Field Unit
  on the James Webb Space Telescope's Near-Infrared Spectrograph}

\bibitem[\protect\citeauthoryear{{Combes}, {Garc{\'{\i}}a-Burillo}, {Boone},
  {Hunt}, {Baker}, {Eckart}, {Englmaier}, {Leon}, {Neri}, {Schinnerer} \&
  {Tacconi}}{{Combes} et~al.}{2004}]{Combes2004}
{Combes} F.,  {Garc{\'{\i}}a-Burillo} S.,  {Boone} F.,  {Hunt} L.~K.,  {Baker}
  A.~J.,  {Eckart} A.,  {Englmaier} P.,  {Leon} S.,  {Neri} R.,  {Schinnerer}
  E.,    {Tacconi} L.~J.,  2004, \aap, 414, 857

\bibitem[\protect\citeauthoryear{{Cox}, {Jonsson}, {Somerville}, {Primack} \&
  {Dekel}}{{Cox} et~al.}{2008}]{Cox2008}
{Cox} T.~J.,  {Jonsson} P.,  {Somerville} R.~S.,  {Primack} J.~R.,    {Dekel}
  A.,  2008, \mnras, 384, 386

\bibitem[\protect\citeauthoryear{{Cutri}, {Rieke}, {Tokunaga}, {Willner} \&
  {Rudy}}{{Cutri} et~al.}{1984}]{Cutri1984}
{Cutri} R.~M.,  {Rieke} G.~H.,  {Tokunaga} A.~T.,  {Willner} S.~P.,    {Rudy}
  R.~J.,  1984, \apj, 280, 521

\bibitem[\protect\citeauthoryear{{de Vaucouleurs}, {de Vaucouleurs} \&
  {Corwin}}{{de Vaucouleurs} et~al.}{1976}]{deVauc1976}
{de Vaucouleurs} G.,  {de Vaucouleurs} A.,    {Corwin} J.~R.,  1976, in Second
  reference catalogue of bright galaxies, 1976, Austin: University of Texas
  Press. {Second reference catalogue of bright galaxies}.
pp~0--+

\bibitem[\protect\citeauthoryear{{D{\'{\i}}az-Santos}, {Alonso-Herrero},
  {Colina}, {Packham}, {Levenson}, {Pereira-Santaella}, {Roche} \&
  {Telesco}}{{D{\'{\i}}az-Santos} et~al.}{2010}]{Diazsantos2010}
{D{\'{\i}}az-Santos} T.,  {Alonso-Herrero} A.,  {Colina} L.,  {Packham} C.,
  {Levenson} N.~A.,  {Pereira-Santaella} M.,  {Roche} P.~F.,    {Telesco}
  C.~M.,  2010, \apj, 711, 328

\bibitem[\protect\citeauthoryear{{D{\'{\i}}az-Santos}, {Alonso-Herrero},
  {Colina}, {Packham}, {Radomski} \& {Telesco}}{{D{\'{\i}}az-Santos}
  et~al.}{2008}]{Diazsantos2008}
{D{\'{\i}}az-Santos} T.,  {Alonso-Herrero} A.,  {Colina} L.,  {Packham} C.,
  {Radomski} J.~T.,    {Telesco} C.~M.,  2008, \apj, 685, 211

\bibitem[\protect\citeauthoryear{{Downes} \& {Solomon}}{{Downes} \&
  {Solomon}}{1998}]{Downes1998}
{Downes} D.,  {Solomon} P.~M.,  1998, \apj, 507, 615

\bibitem[\protect\citeauthoryear{{Eliche-Moral}, {Gonz{\'a}lez-Garc{\'{\i}}a},
  {Balcells}, {Aguerri}, {Gallego}, {Zamorano} \& {Prieto}}{{Eliche-Moral}
  et~al.}{2011}]{ElicheMoral2011}
{Eliche-Moral} M.~C.,  {Gonz{\'a}lez-Garc{\'{\i}}a} A.~C.,  {Balcells} M.,
  {Aguerri} J.~A.~L.,  {Gallego} J.,  {Zamorano} J.,    {Prieto} M.,  2011,
  ArXiv e-prints

\bibitem[\protect\citeauthoryear{{Farrah}, {Bernard-Salas}, {Spoon}, {Soifer},
  {Armus}, {Brandl}, {Charmandaris}, {Desai}, {Higdon}, {Devost} \&
  {Houck}}{{Farrah} et~al.}{2007}]{Farrah2007}
{Farrah} D.,  {Bernard-Salas} J.,  {Spoon} H.~W.~W.,  {Soifer} B.~T.,  {Armus}
  L.,  {Brandl} B.,  {Charmandaris} V.,  {Desai} V.,  {Higdon} S.,  {Devost}
  D.,    {Houck} J.,  2007, \apj, 667, 149

\bibitem[\protect\citeauthoryear{{Farrah}, {Rowan-Robinson}, {Oliver},
  {Serjeant}, {Borne}, {Lawrence}, {Lucas}, {Bushouse} \& {Colina}}{{Farrah}
  et~al.}{2001}]{Farrah2001}
{Farrah} D.,  {Rowan-Robinson} M.,  {Oliver} S.,  {Serjeant} S.,  {Borne} K.,
  {Lawrence} A.,  {Lucas} R.~A.,  {Bushouse} H.,    {Colina} L.,  2001, \mnras,
  326, 1333

\bibitem[\protect\citeauthoryear{{Galliano}, {Madden}, {Tielens}, {Peeters} \&
  {Jones}}{{Galliano} et~al.}{2008}]{Galliano2008}
{Galliano} F.,  {Madden} S.~C.,  {Tielens} A.~G.~G.~M.,  {Peeters} E.,
  {Jones} A.~P.,  2008, \apj, 679, 310

\bibitem[\protect\citeauthoryear{{Haan}, {Surace}, {Armus}, {Evans}, {Howell},
  {Mazzarella}, {Kim}, {Vavilkin}, {Inami}, {Sanders}, {Petric}, {Bridge},
  {Melbourne}, {Charmandaris}, {Diaz-Santos}, {Murphy}, {U}, {Stierwalt} \&
  {Marshall}}{{Haan} et~al.}{2011}]{Haan2011}
{Haan} S.,  {Surace} J.~A.,  {Armus} L.,  {Evans} A.~S.,  {Howell} J.~H.,
  {Mazzarella} J.~M.,  {Kim} D.~C.,  {Vavilkin} T.,  {Inami} H.,  {Sanders}
  D.~B.,  {Petric} A.,  {Bridge} C.~R.,  {Melbourne} J.~L.,  {Charmandaris} V.,
   {Diaz-Santos} T.,  {Murphy} E.~J.,  {U} V.,  {Stierwalt} S.,    {Marshall}
  J.~A.,  2011, \aj, 141, 100

\bibitem[\protect\citeauthoryear{{Heisler}, {Dopita}, {Kewley} \&
  {Lumsden}}{{Heisler} et~al.}{1999}]{Heisler1999}
{Heisler} C.~A.,  {Dopita} M.~A.,  {Kewley} L.~J.,    {Lumsden} S.~L.,  1999,
  \apss, 266, 181

\bibitem[\protect\citeauthoryear{{Hopkins}, {Hernquist}, {Cox}, {Di Matteo},
  {Robertson} \& {Springel}}{{Hopkins} et~al.}{2006}]{Hopkins2006}
{Hopkins} P.~F.,  {Hernquist} L.,  {Cox} T.~J.,  {Di Matteo} T.,  {Robertson}
  B.,    {Springel} V.,  2006, \apjs, 163, 1

\bibitem[\protect\citeauthoryear{{Hsieh}, {Matsushita}, {Liu}, {Ho}, {Oi} \&
  {Wu}}{{Hsieh} et~al.}{2011}]{Hsieh2011}
{Hsieh} P.-Y.,  {Matsushita} S.,  {Liu} G.,  {Ho} P.~T.~P.,  {Oi} N.,    {Wu}
  Y.-L.,  2011, \apj, 736, 129

\bibitem[\protect\citeauthoryear{{Imanishi} \& {Dudley}}{{Imanishi} \&
  {Dudley}}{2000}]{Imanishi2000}
{Imanishi} M.,  {Dudley} C.~C.,  2000, \apj, 545, 701

\bibitem[\protect\citeauthoryear{{Imanishi}, {Dudley} \& {Maloney}}{{Imanishi}
  et~al.}{2006}]{Imanishi2006}
{Imanishi} M.,  {Dudley} C.~C.,    {Maloney} P.~R.,  2006, \apj, 637, 114

\bibitem[\protect\citeauthoryear{{Imanishi}, {Imase}, {Oi} \&
  {Ichikawa}}{{Imanishi} et~al.}{2011}]{Imanishi2011}
{Imanishi} M.,  {Imase} K.,  {Oi} N.,    {Ichikawa} K.,  2011, \aj, 141, 156

\bibitem[\protect\citeauthoryear{{Imanishi}, {Nakagawa}, {Shirahata}, {Ohyama}
  \& {Onaka}}{{Imanishi} et~al.}{2010}]{Imanishi2010}
{Imanishi} M.,  {Nakagawa} T.,  {Shirahata} M.,  {Ohyama} Y.,    {Onaka} T.,
  2010, \apj, 721, 1233

\bibitem[\protect\citeauthoryear{{Johansson}, {Burkert} \& {Naab}}{{Johansson}
  et~al.}{2009a}]{Peter2009a}
{Johansson} P.~H.,  {Burkert} A.,    {Naab} T.,  2009a, \apjl, 707, L184

\bibitem[\protect\citeauthoryear{{Johansson}, {Naab} \& {Burkert}}{{Johansson}
  et~al.}{2009b}]{Johansson2009b}
{Johansson} P.~H.,  {Naab} T.,    {Burkert} A.,  2009b, \apj, 690, 802

\bibitem[\protect\citeauthoryear{{Keller}, {Sloan}, {Forrest}, {Ayala},
  {D'Alessio}, {Shah}, {Calvet}, {Najita}, {Li}, {Hartmann}, {Sargent},
  {Watson} \& {Chen}}{{Keller} et~al.}{2008}]{Keller2008}
{Keller} L.~D.,  {Sloan} G.~C.,  {Forrest} W.~J.,  {Ayala} S.,  {D'Alessio} P.,
   {Shah} S.,  {Calvet} N.,  {Najita} J.,  {Li} A.,  {Hartmann} L.,  {Sargent}
  B.,  {Watson} D.~M.,    {Chen} C.~H.,  2008, \apj, 684, 411

\bibitem[\protect\citeauthoryear{{Kendrew}, {Jolissaint}, {Brandl}, {Lenzen},
  {Pantin}, {Glasse}, {Blommaert}, {Venema}, {Siebenmorgen} \&
  {Molster}}{{Kendrew} et~al.}{2010}]{Kendrew2010}
{Kendrew} S.,  {Jolissaint} L.,  {Brandl} B.,  {Lenzen} R.,  {Pantin} E.,
  {Glasse} A.,  {Blommaert} J.,  {Venema} L.,  {Siebenmorgen} R.,    {Molster}
  F.,  2010, in Society of Photo-Optical Instrumentation Engineers (SPIE)
  Conference Series Vol.~7735 of Society of Photo-Optical Instrumentation
  Engineers (SPIE) Conference Series, {Mid-infrared astronomy with the E-ELT:
  performance of METIS}

\bibitem[\protect\citeauthoryear{{Keto}, {Ball}, {Arens}, {Jernigan} \&
  {Meixner}}{{Keto} et~al.}{1992}]{Keto1992}
{Keto} E.,  {Ball} R.,  {Arens} J.,  {Jernigan} G.,    {Meixner} M.,  1992,
  \apj, 389, 223

\bibitem[\protect\citeauthoryear{{Kotilainen}, {Reunanen}, {Laine} \&
  {Ryder}}{{Kotilainen} et~al.}{2000}]{Kotilainen2000}
{Kotilainen} J.~K.,  {Reunanen} J.,  {Laine} S.,    {Ryder} S.~D.,  2000, \aap,
  353, 834

\bibitem[\protect\citeauthoryear{{Kotilainen}, {Reunanen}, {Laine} \&
  {Ryder}}{{Kotilainen} et~al.}{2001}]{Kotilainen2001}
{Kotilainen} J.~K.,  {Reunanen} J.,  {Laine} S.,    {Ryder} S.~D.,  2001, \aap,
  366, 439

\bibitem[\protect\citeauthoryear{{Laine}, {Kotilainen}, {Reunanen}, {Ryder} \&
  {Beck}}{{Laine} et~al.}{2006}]{Laine2006}
{Laine} S.,  {Kotilainen} J.~K.,  {Reunanen} J.,  {Ryder} S.~D.,    {Beck} R.,
  2006, \aj, 131, 701

\bibitem[\protect\citeauthoryear{{Lebouteiller}, {Bernard-Salas}, {Whelan},
  {Brandl}, {Galliano}, {Charmandaris}, {Madden} \& {Kunth}}{{Lebouteiller}
  et~al.}{2011}]{Lebouteiller2011}
{Lebouteiller} V.,  {Bernard-Salas} J.,  {Whelan} D.~G.,  {Brandl} B.,
  {Galliano} F.,  {Charmandaris} V.,  {Madden} S.,    {Kunth} D.,  2011, \apj,
  728, 45

\bibitem[\protect\citeauthoryear{{Leitherer}, {Schaerer}, {Goldader},
  {Gonz{\'a}lez Delgado}, {Robert}, {Kune}, {de Mello}, {Devost} \&
  {Heckman}}{{Leitherer} et~al.}{1999}]{Leitherer1999}
{Leitherer} C.,  {Schaerer} D.,  {Goldader} J.~D.,  {Gonz{\'a}lez Delgado}
  R.~M.,  {Robert} C.,  {Kune} D.~F.,  {de Mello} D.~F.,  {Devost} D.,
  {Heckman} T.~M.,  1999, \apjs, 123, 3

\bibitem[\protect\citeauthoryear{{Levenson}, {Sirocky}, {Hao}, {Spoon},
  {Marshall}, {Elitzur} \& {Houck}}{{Levenson} et~al.}{2007}]{Levenson2007}
{Levenson} N.~A.,  {Sirocky} M.~M.,  {Hao} L.,  {Spoon} H.~W.~W.,  {Marshall}
  J.~A.,  {Elitzur} M.,    {Houck} J.~R.,  2007, \apjl, 654, L45

\bibitem[\protect\citeauthoryear{{Lonsdale}, {Farrah} \& {Smith}}{{Lonsdale}
  et~al.}{2006}]{Lonsdale2006}
{Lonsdale} C.~J.,  {Farrah} D.,    {Smith} H.~E.,  2006, in {Mason, J.~W.} ed.,
  Astrophysics Update 2 {Ultraluminous Infrared Galaxies}.
Springer Verlag, pp 285--+

\bibitem[\protect\citeauthoryear{{Mason}, {Levenson}, {Packham}, {Elitzur},
  {Radomski}, {Petric} \& {Wright}}{{Mason} et~al.}{2007}]{Mason2007}
{Mason} R.~E.,  {Levenson} N.~A.,  {Packham} C.,  {Elitzur} M.,  {Radomski} J.,
   {Petric} A.~O.,    {Wright} G.~S.,  2007, \apj, 659, 241

\bibitem[\protect\citeauthoryear{{Mazzarella}, {Voit}, {Soifer}, {Matthews},
  {Graham}, {Armus} \& {Shupe}}{{Mazzarella} et~al.}{1994}]{Mazzarella1994}
{Mazzarella} J.~M.,  {Voit} G.~M.,  {Soifer} B.~T.,  {Matthews} K.,  {Graham}
  J.~R.,  {Armus} L.,    {Shupe} D.,  1994, \aj, 107, 1274

\bibitem[\protect\citeauthoryear{{Mihos} \& {Hernquist}}{{Mihos} \&
  {Hernquist}}{1994}]{Mihos1994}
{Mihos} J.~C.,  {Hernquist} L.,  1994, \apjl, 425, L13

\bibitem[\protect\citeauthoryear{{Mizutani}, {Suto} \& {Maihara}}{{Mizutani}
  et~al.}{1994}]{Mizutani1994}
{Mizutani} K.,  {Suto} H.,    {Maihara} T.,  1994, \apj, 421, 475

\bibitem[\protect\citeauthoryear{{Murphy} Jr., {Armus}, {Matthews}, {Soifer},
  {Mazzarella}, {Shupe}, {Strauss} \& {Neugebauer}}{{Murphy}
  et~al.}{1996}]{Murphy1996}
{Murphy} Jr. T.~W.,  {Armus} L.,  {Matthews} K.,  {Soifer} B.~T.,  {Mazzarella}
  J.~M.,  {Shupe} D.~L.,  {Strauss} M.~A.,    {Neugebauer} G.,  1996, \aj, 111,
  1025

\bibitem[\protect\citeauthoryear{{Murphy} Jr., {Soifer}, {Matthews} \&
  {Armus}}{{Murphy} et~al.}{2001}]{Murphy2001}
{Murphy} Jr. T.~W.,  {Soifer} B.~T.,  {Matthews} K.,    {Armus} L.,  2001,
  \apj, 559, 201

\bibitem[\protect\citeauthoryear{{Neff}, {Hutchings}, {Standord} \&
  {Unger}}{{Neff} et~al.}{1990}]{Neff1990}
{Neff} S.~G.,  {Hutchings} J.~B.,  {Standord} S.~A.,    {Unger} S.~W.,  1990,
  \aj, 99, 1088

\bibitem[\protect\citeauthoryear{{Oi}, {Imanishi} \& {Imase}}{{Oi}
  et~al.}{2010}]{Oi2010}
{Oi} N.,  {Imanishi} M.,    {Imase} K.,  2010, PASJ, 62, 1509

\bibitem[\protect\citeauthoryear{{Olsson}, {Aalto}, {Thomasson} \&
  {Beswick}}{{Olsson} et~al.}{2010}]{Olsson2010}
{Olsson} E.,  {Aalto} S.,  {Thomasson} M.,    {Beswick} R.,  2010, \aap, 513,
  A11+

\bibitem[\protect\citeauthoryear{{Peeters}, {Spoon} \& {Tielens}}{{Peeters}
  et~al.}{2004}]{Peeters2004}
{Peeters} E.,  {Spoon} H.~W.~W.,    {Tielens} A.~G.~G.~M.,  2004, \apj, 613,
  986

\bibitem[\protect\citeauthoryear{{Puxley} \& {Brand}}{{Puxley} \&
  {Brand}}{1999}]{Puxley1999}
{Puxley} P.~J.,  {Brand} P.~W.~J.~L.,  1999, \apj, 514, 675

\bibitem[\protect\citeauthoryear{{Ramsay Howat}, {Todd}, {Leggett}, {Davis},
  {Strachan}, {Borrowman}, {Ellis}, {Elliot}, {Gostick}, {Kackley} \&
  {Rippa}}{{Ramsay Howat} et~al.}{2004}]{Ramsay2004}
{Ramsay Howat} S.~K.,  {Todd} S.,  {Leggett} S.,  {Davis} C.,  {Strachan} M.,
  {Borrowman} A.,  {Ellis} M.,  {Elliot} J.,  {Gostick} D.,  {Kackley} R.,
  {Rippa} M.,  2004, in {A.~F.~M.~Moorwood \& M.~Iye} ed., Society of
  Photo-Optical Instrumentation Engineers (SPIE) Conference Series Vol.~5492 of
  Presented at the Society of Photo-Optical Instrumentation Engineers (SPIE)
  Conference, {The commissioning of and first results from the UIST imager
  spectrometer}.
pp 1160--1171

\bibitem[\protect\citeauthoryear{{Reunanen}, {Prieto} \&
  {Siebenmorgen}}{{Reunanen} et~al.}{2010}]{Reunanen2010}
{Reunanen} J.,  {Prieto} M.~A.,    {Siebenmorgen} R.,  2010, \mnras, 402, 879

\bibitem[\protect\citeauthoryear{{Reunanen}, {Tacconi-Garman} \&
  {Ivanov}}{{Reunanen} et~al.}{2007}]{Reunanen2007}
{Reunanen} J.,  {Tacconi-Garman} L.~E.,    {Ivanov} V.~D.,  2007, \mnras, 382,
  951

\bibitem[\protect\citeauthoryear{{Risaliti}, {Gilli}, {Maiolino} \&
  {Salvati}}{{Risaliti} et~al.}{2000}]{Risaliti2000}
{Risaliti} G.,  {Gilli} R.,  {Maiolino} R.,    {Salvati} M.,  2000, \aap, 357,
  13

\bibitem[\protect\citeauthoryear{{Risaliti}, {Imanishi} \& {Sani}}{{Risaliti}
  et~al.}{2010}]{Risaliti2010}
{Risaliti} G.,  {Imanishi} M.,    {Sani} E.,  2010, \mnras, 401, 197

\bibitem[\protect\citeauthoryear{{Risaliti}, {Maiolino}, {Marconi}, {Sani},
  {Berta}, {Braito}, {Della Ceca}, {Franceschini} \& {Salvati}}{{Risaliti}
  et~al.}{2006}]{Risaliti2006}
{Risaliti} G.,  {Maiolino} R.,  {Marconi} A.,  {Sani} E.,  {Berta} S.,
  {Braito} V.,  {Della Ceca} R.,  {Franceschini} A.,    {Salvati} M.,  2006,
  \mnras, 365, 303

\bibitem[\protect\citeauthoryear{{Rossa}, {Laine}, {van der Marel}, {Mihos},
  {Hibbard}, {B{\"o}ker} \& {Zabludoff}}{{Rossa} et~al.}{2007}]{Rossa2007}
{Rossa} J.,  {Laine} S.,  {van der Marel} R.~P.,  {Mihos} J.~C.,  {Hibbard}
  J.~E.,  {B{\"o}ker} T.,    {Zabludoff} A.~I.,  2007, \aj, 134, 2124

\bibitem[\protect\citeauthoryear{{Rossa}, {van der Marel}, {B{\"o}ker},
  {Gerssen}, {Ho}, {Rix}, {Shields} \& {Walcher}}{{Rossa}
  et~al.}{2006}]{Rossa2006}
{Rossa} J.,  {van der Marel} R.~P.,  {B{\"o}ker} T.,  {Gerssen} J.,  {Ho}
  L.~C.,  {Rix} H.,  {Shields} J.~C.,    {Walcher} C.,  2006, \aj, 132, 1074

\bibitem[\protect\citeauthoryear{{Sanders}, {Soifer}, {Elias}, {Madore},
  {Matthews}, {Neugebauer} \& {Scoville}}{{Sanders} et~al.}{1988}]{Sanders1988}
{Sanders} D.~B.,  {Soifer} B.~T.,  {Elias} J.~H.,  {Madore} B.~F.,  {Matthews}
  K.,  {Neugebauer} G.,    {Scoville} N.~Z.,  1988, \apj, 325, 74

\bibitem[\protect\citeauthoryear{{Sani}, {Risaliti}, {Salvati}, {Maiolino},
  {Marconi}, {Berta}, {Braito}, {Della Ceca} \& {Franceschini}}{{Sani}
  et~al.}{2008}]{Sani2008}
{Sani} E.,  {Risaliti} G.,  {Salvati} M.,  {Maiolino} R.,  {Marconi} A.,
  {Berta} S.,  {Braito} V.,  {Della Ceca} R.,    {Franceschini} A.,  2008,
  \apj, 675, 96

\bibitem[\protect\citeauthoryear{{Schwartz} \& {Martin}}{{Schwartz} \&
  {Martin}}{2004}]{Schwartz2004}
{Schwartz} C.~M.,  {Martin} C.~L.,  2004, \apj, 610, 201

\bibitem[\protect\citeauthoryear{{Siebenmorgen}, {Haas}, {Pantin},
  {Kr{\"u}gel}, {Leipski}, {K{\"a}ufl}, {Lagage}, {Moorwood}, {Smette} \&
  {Sterzik}}{{Siebenmorgen} et~al.}{2008}]{Siebenmorgen2008}
{Siebenmorgen} R.,  {Haas} M.,  {Pantin} E.,  {Kr{\"u}gel} E.,  {Leipski} C.,
  {K{\"a}ufl} H.~U.,  {Lagage} P.~O.,  {Moorwood} A.,  {Smette} A.,
  {Sterzik} M.,  2008, \aap, 488, 83

\bibitem[\protect\citeauthoryear{{Soifer}, {Neugebauer}, {Matthews}, {Egami},
  {Weinberger}, {Ressler}, {Scoville}, {Stolovy}, {Condon} \&
  {Becklin}}{{Soifer} et~al.}{2001}]{Soifer2001}
{Soifer} B.~T.,  {Neugebauer} G.,  {Matthews} K.,  {Egami} E.,  {Weinberger}
  A.~J.,  {Ressler} M.,  {Scoville} N.~Z.,  {Stolovy} S.~R.,  {Condon} J.~J.,
   {Becklin} E.~E.,  2001, \aj, 122, 1213

\bibitem[\protect\citeauthoryear{{Spoon}, {Marshall}, {Houck}, {Elitzur},
  {Hao}, {Armus}, {Brandl} \& {Charmandaris}}{{Spoon} et~al.}{2007}]{Spoon2007}
{Spoon} H.~W.~W.,  {Marshall} J.~A.,  {Houck} J.~R.,  {Elitzur} M.,  {Hao} L.,
  {Armus} L.,  {Brandl} B.~R.,    {Charmandaris} V.,  2007, \apjl, 654, L49

\bibitem[\protect\citeauthoryear{{Springel}, {Di Matteo} \&
  {Hernquist}}{{Springel} et~al.}{2005}]{Springel2005}
{Springel} V.,  {Di Matteo} T.,    {Hernquist} L.,  2005, \apjl, 620, L79

\bibitem[\protect\citeauthoryear{{Surace}, {Sanders} \& {Evans}}{{Surace}
  et~al.}{2000}]{Surace2000}
{Surace} J.~A.,  {Sanders} D.~B.,    {Evans} A.~S.,  2000, \apj, 529, 170

\bibitem[\protect\citeauthoryear{{Tacconi-Garman}, {Sturm}, {Lehnert}, {Lutz},
  {Davies} \& {Moorwood}}{{Tacconi-Garman} et~al.}{2005}]{Tacconi2005}
{Tacconi-Garman} L.~E.,  {Sturm} E.,  {Lehnert} M.,  {Lutz} D.,  {Davies}
  R.~I.,    {Moorwood} A.~F.~M.,  2005, \aap, 432, 91

\bibitem[\protect\citeauthoryear{{Tielens}}{{Tielens}}{2008}]{Tielens2008}
{Tielens} A.~G.~G.~M.,  2008, \araa, 46, 289

\bibitem[\protect\citeauthoryear{{V{\"a}is{\"a}nen}, {Mattila}, {Kniazev},
  {Adamo}, {Efstathiou}, {Farrah}, {Johansson}, {Still} \&
  {Zijlstra}}{{V{\"a}is{\"a}nen} et~al.}{2008}]{Vaisanen2008}
{V{\"a}is{\"a}nen} P.,  {Mattila} S.,  {Kniazev} A.,  {Adamo} A.,  {Efstathiou}
  A.,  {Farrah} D.,  {Johansson} P.~H.,  {Still} M.,    {Zijlstra} A.,  2008,
  \mnras, 384, 886

\bibitem[\protect\citeauthoryear{{Veilleux}, {Kim}, {Sanders}, {Mazzarella} \&
  {Soifer}}{{Veilleux} et~al.}{1995}]{Veilleux1995}
{Veilleux} S.,  {Kim} D.-C.,  {Sanders} D.~B.,  {Mazzarella} J.~M.,    {Soifer}
  B.~T.,  1995, \apjs, 98, 171

\bibitem[\protect\citeauthoryear{{Voit}}{{Voit}}{1992}]{Voit1992}
{Voit} G.~M.,  1992, \mnras, 258, 841

\bibitem[\protect\citeauthoryear{{Watabe}, {Kawakatu} \& {Imanishi}}{{Watabe}
  et~al.}{2008}]{Watabe2008}
{Watabe} Y.,  {Kawakatu} N.,    {Imanishi} M.,  2008, \apj, 677, 895

\bibitem[\protect\citeauthoryear{{Yuan}, {Kewley} \& {Sanders}}{{Yuan}
  et~al.}{2010}]{Yuan2010}
{Yuan} T.-T.,  {Kewley} L.~J.,    {Sanders} D.~B.,  2010, \apj, 709, 884

\bibitem[\protect\citeauthoryear{{Zakamska}}{{Zakamska}}{2010}]{Zakamska2010}
{Zakamska} N.~L.,  2010, \nat, 465, 60

\end{thebibliography}

\end{document}